\documentclass[preprint,12pt]{emulateapj}
\usepackage{natbib} 
\usepackage{times}
\usepackage{epsfig,graphicx}
\usepackage{txfonts}

\newcommand{\Msun}{$M_{\odot}$}

\newcommand{\kms}{km s$^{-1}$}
\newcommand{\h}{^h}
\newcommand{\m}{^m}
\newcommand{\s}{^s}
\newcommand{\dg}{^\circ}

\newcommand{\am}{'}

\newcommand{\as}{''}
\newcommand{\cprops}{$\it{cprops}$\space}
\newcommand{\cpropse}{$\it{cprops}$}

\newcommand{\xco}{$X_{\text{CO}}$~}
\newcommand{\xcons}{$X_{\text{CO}}$}
\newcommand{\xcounits}{$\times$ 10$^{20}$ cm$^{-2}$ / (K km s$^{-1}$)}

\begin{document}

\title{Resolved giant molecular clouds in nearby spiral galaxies: \\ insights from the CANON CO (1-0) Survey \\  \textnormal{\today}}
\author{Jennifer Donovan Meyer\altaffilmark{1,2}, Jin Koda\altaffilmark{1}, Rieko Momose\altaffilmark{3,4,5}, Thomas Mooney\altaffilmark{1}, Fumi Egusa\altaffilmark{6,7}, Misty Carty\altaffilmark{8}, Robert Kennicutt\altaffilmark{9}, Nario Kuno\altaffilmark{10,11},  David Rebolledo\altaffilmark{12}, Tsuyoshi Sawada\altaffilmark{4,13}, Nick Scoville\altaffilmark{7}, Tony Wong\altaffilmark{12} }
\altaffiltext{1}{Department of Physics \& Astronomy, Stony Brook University, Stony Brook, NY 11794}
\altaffiltext{2}{National Radio Astronomy Observatory, Charlottesville, VA 22901}
\altaffiltext{3}{Department of Astronomy, University of Tokyo, Hongo, Bunkyo-ku, Tokyo 113-0033, Japan}
\altaffiltext{4}{National Astronomical Observatory of Japan, Mitaka, Tokyo 181-8588, Japan}
\altaffiltext{5}{Institute for Cosmic Ray Research, University of Tokyo, 5-1-5 Kashiwa-no-Ha, Kashiwa City, Chiba, 277-8582, Japan}
\altaffiltext{6}{Institute of Space and Astronautical Science, Japan Aerospace Exploration Agency, Chuo-ku, Sagamihara, Kanagawa 252-5210, Japan}
\altaffiltext{7}{Department of Astronomy, California Institute of Technology, Pasadena, CA 91125}
\altaffiltext{8}{Department of Astronomy, University of Maryland, College Park, MD 20742}
\altaffiltext{9}{Institute of Astronomy, University of Cambridge, Cambridge CB3 0HA, United Kingdom}
\altaffiltext{10}{Nobeyama Radio Observatory, Minamimaki, Minamisaku, Nagano, 384-1305, Japan}
\altaffiltext{11}{The Graduate University for Advanced Studies (SOKENDAI), 2-21-1 Osawa, Mitaka, Tokyo 181-0015}
\altaffiltext{12}{Astronomy Department, University of Illinois, Urbana, IL 61801}
\altaffiltext{13}{Joint ALMA Observatory, Alonso de C\'ordova 3107, Vitacura, Santiago 763-0355, Chile}

\begin{abstract}
We resolve 182 individual giant molecular clouds (GMCs) larger than 2.5 $\times$ 10$^{5}$ \Msun~in the inner disks of five large nearby spiral galaxies (NGC 2403, NGC 3031, NGC 4736, NGC 4826, and NGC 6946) to create the largest such sample of extragalactic GMCs within galaxies analogous to the Milky Way. Using a conservatively chosen sample of GMCs most likely to adhere to the virial assumption, we measure cloud sizes, velocity dispersions, and $^{12}$CO (J=1-0) luminosities and calculate cloud virial masses. The average conversion factor from CO flux to H$_{2}$ mass (or \xcons) for each galaxy is 1-2 \xcounits, all within a factor of two of the Milky Way disk value ($\sim$2 \xcounits). We find GMCs to be generally consistent within our errors between the galaxies and with Milky Way disk GMCs; the intrinsic scatter between clouds is of order a factor of two. Consistent with previous studies in the Local Group, we find a linear relationship between cloud virial mass and CO luminosity, supporting the assumption that the clouds in this GMC sample are gravitationally bound. We do not detect a significant population of GMCs with elevated velocity dispersions for their sizes, as has been detected in the Galactic center. Though the range of metallicities probed in this study is narrow, the average conversion factors of these galaxies will serve to anchor the high metallicity end of metallicity-\xco trends measured using conversion factors in resolved clouds; this has been previously possible primarily with Milky Way measurements. 
\end{abstract}


\section{Introduction}

Measuring the amount of gas available for star formation in galaxies at all redshifts is crucial to understanding how galaxies evolve from the early universe to the present day. Most of the mass of gas in galaxies is composed of atomic and molecular hydrogen, with hydrogen gas in the latter phase generally believed to be the direct fuel for star formation. However, as a result of the cold temperatures in molecular gas environments, H$_{2}$ line emission cannot be directly observed in the disks of most galaxies, and other molecules are typically used to trace H$_{2}$ mass. The most common of these tracers are the low rotational transitions of CO -- and in particular, the J=1-0 transition -- which are easily observable from the ground; at temperatures typical of bulk GMC gas, optically thick $^{12}$CO (J=1-0), with an energy of 5.5~K above ground, is collisionally excited at densities of only a few hundred particles per cm$^{-3}$. Determining the conversion factor between CO flux and H$_{2}$ mass, called \xco or $\alpha_{\text{CO}}$ in the literature, therefore has important implications for studies of galaxies throughout cosmic history. 

In the Milky Way disk, the conversion factor is fairly well constrained. When individual GMCs can be resolved, comparing CO luminosities and virial masses of individual clouds effectively allows a direct measure of the mass-to-light ratio in each GMC. Such virial mass-based measurements of \xco in the Milky Way disk indicate that the value is within a factor of two of 2-3 \xcounits (\citealt{Scoville87, Solomon87}, hereafter S87). Predicting the locations of expected CO emission using far infrared and HI maps of the Galaxy and comparing them to actual Galactic CO yields a value of 1.8 $\pm$ 0.3 \xcounits~ \citep{Dame01}. The Milky Way value of \xco has also been constrained using a method in which the mass is traced using gamma ray observations; the results of these studies are within the errors of the \citet{Dame01} result \citep{Strong96, Hunter97}. \citet{McKee07} recalculate the S87 value of the conversion factor for a 10$^{6}$ \Msun~GMC using an updated Galactic center distance and find a value of 1.9 \xcounits, consistent with the other non-virial measurements cited here (supporting the S87 claim that GMCs are bound on average).

To date, due to limitations in instrumental resolution, the majority of extragalactic GMC virial mass measurements have been in nearby dwarfs and small spirals. \citet{Wilson90} measured the first extragalactic conversion factor within the GMCs of M33, finding it to be consistent with the Galactic value. In the NANTEN surveys of the Large Magellanic Cloud (LMC) \citep{Mizuno01, Fukui08}, average conversion factors 3-4 times higher than the Galactic value -- and a value at least ten times the Galactic value in the Small Magellanic Cloud (SMC) -- are found using virial measurements \citep{Mizuno01_SMC}. Conversion factors higher than the Milky Way value have been observed in nearby dwarf and irregular galaxies using individual, resolved GMCs \citep{Dettmar89, Wilson91, Gratier2010, Hughes2010} and in the SMC using a range of methods \citep{Rubio91, Rubio93, Israel97, Leroy07, Blitz07, Bolatto08}.

The NANTEN survey data and other complete CO studies of nearby galaxies (including observations of IC10 [\citealt{Leroy06}], M33 [\citealt{Engargiola03}], and one of the spiral arms of M31 [\citealt{Rosolowsky07}]) are summarized and uniformly analyzed in \citet{Blitz07}, who find 4 \xcounits~to be the average conversion factor for the Local Group; this value also yields dust-to-gas ratios consistent with Milky Way observations and predictions of the dust models of the SINGS galaxies presented in \citet{Draine07}. The Local Group sample is extended in Bolatto et al. (2008, hereafter B08) with the inclusion of even more dwarfs (B08 and references therein), where virial measurements of extragalactic GMCs are found to be largely consistent with Galactic GMCs. 

The ability to resolve individual GMCs limits the distance to which \xco can be reliably measured using the virial mass method, but conversion factors in resolved GMCs have also been derived in more substantial nearby galaxies. \citet{Nakai95} derive an average conversion factor toward HII regions in the spiral arms of M51 from measurements of visual extinction and CO intensities of 1.6 \xcounits~(rescaled using the more recent calibration for the CO flux from \citealt{Koda11}), with this value increasing in the outer regions of the galaxy. \citet{Adler92} also derive an average value for the conversion factor that is below the Galactic value (1.2 \xcounits) in two regions centered on the spiral arms of M51. \citet{Tosaki03} find GMCs with properties similar to Milky Way GMCs (within a factor of two of the Galactic \xcons) in the outer region of the flocculent galaxy NGC 5055.
\citet{DonovanMeyer12} (hereafter DM12) present a sample of $^{12}$CO (J=1-0) GMCs in NGC 6946 and find that the average conversion factor in the central kiloparsecs is 1.2 \xcounits, again within a factor of two of the Galactic value, which is consistent with the results of \citet{Rebolledo12} in the outer disk of NGC 6946. For a more comprehensive review of studies of the conversion factor, see \citet{Fukui2010}. 

Measuring individual GMC properties in various environments -- from starburst environments (e.g., \citealt{Tacconi08, Genzel12}) to galactic centers (e.g., \citealt{Oka01} in the Milky Way) to dwarf galaxies (e.g., B08) -- lends insight into the physical nature of these clouds over the full range of parameter space, which may ultimately reveal the physical processes that govern the conversion factor. For example, variations in \xco have long been suspected to be due to changes in metallicity such that more metal-poor gas yields a higher conversion factor \citep{Wilson95, Arimoto96, Israel97, Israel00, Leroy11, Feldmann2012}. Generally, as the metallicity decreases, the H$_{2}$ molecule formation rate decreases since it is thought to form on dust grains \citep{Hollenbach71, Genzel92}. In addition, CO is more easily photodissociated due to subsequently poorer H$_{2}$ shielding, leading to a smaller CO luminosity per unit gas mass, effectively raising the conversion factor. \citet{Israel97} also suggests that increased radiation field energy densities induce the same effect. Models of individual GMCs and the effects that metallicity, radiation fields, and cloud mass surface densities (among other physical properties of the ISM) have on the conversion factor are now beginning to surface (i.e., \citealt{Feldmann2012, Narayanan12}, and references therein; \citealt{NarayananHopkins2012}).

In this paper, we present the molecular content of the five best resolved galaxies in our CANON (CArma and NObeyama Nearby galaxies) CO (1-0) Survey (Koda et al., in preparation). The biggest strength of this survey is the presentation of accurate flux measurements yielded by the combination of interferometric and single dish information \citep{Koda09, Koda11}. Each map in this study has a beam size of 78~pc or less, enabling us to identify the largest individual GMCs within our sample galaxies. As we are interested in the physical nature of the interstellar medium contained within GMCs, we calculate individual GMC conversion factors, discuss the gravitational stability of the GMCs, and place the derived conversion factors within the context of the galaxy metallicities.

\section{Observations}
\subsection{CANON CO (1-0) Survey}
The CO (J=1-0) observations presented in this paper were taken as part of the CANON CO (1-0) Survey, in which data from the Combined Array for Research in Millimeter Astronomy (CARMA) and Nobeyama Radio Observatory 45-meter (NRO45) single dish telescope are combined to image the central regions of nearby spiral galaxies (Koda et al., in preparation). The goal of the survey is to study the molecular ISM within nearby spirals with a range of morphologies in order to understand molecular cloud evolution and resolved star formation. The five galaxies presented in this paper are the closest survey galaxies, with distances less than 7.5~Mpc, for which we are able to attain the highest resolutions (less than 78~pc). Their basic properties are listed in Table~\ref{basic}.

\begin{table*}[t]
\begin{center}
\caption{Galaxy properties}
\smallskip
\begin{tabular}{cccccc}
\hline
\hline
& NGC 2403 & NGC 3031 & NGC 4736 & NGC 4826 & NGC 6946 \\
\hline
RA$_{J2000}$ & 07$\h$36$\m$51.4$\s$ & 09$\h$55$\m$33.2$\s$ & 12$\h$50$\m$53.0$\s$ & 12$\h$56$\m$43.7$\s$ & 20$\h$34$\m$52.3$\s$ \\
Dec$_{J2000}$ & +65$\dg$36$\am$09$\as$ & +69$\dg$03$\am$55$\as$ & +41$\dg$07$\am$14$\as$ & +21$\dg$40$\am$52$\as$ & +60$\dg$09$\am$14$\as$ \\
Optical velocity (\kms) & 131 & -34 & 308 & 408 & 48 \\
Distance (Mpc) & 3.13 $\pm$ 0.14 & 3.55 $\pm$ 0.13 & 5.20 $\pm$ 0.43 & 7.48 $\pm$ 0.69 & 6.8 $\pm$ 1.7 \\
Morphology & SABcd & SAab & SAab & SAab & SABcd \\
Inclination angle (degrees) & 57 & 60 & 36 & 59 & 32 \\
Central metallicity & 8.92 $\pm$ 0.02 & 9.15 $\pm$ 0.03 & 9.06 $\pm$ 0.03$^{a}$ & 9.20 $\pm$ 0.04 & 9.16 $\pm$ 0.06 \\
Metallicity gradient (dex $\rho_{25}^{-1}$) & -0.26 & -0.45 & -0.11 & ... & -0.28 \\
$\rho_{25}$ (arcmin) & 10.94 & 13.46 & 5.61 & 5.00 & 5.74 \\
\hline
\end{tabular} \\
\vspace{0.1in}
RA, Dec, and optical velocities are from NED. Distances, inclination angles, metallicities, and $\rho_{25}$ are from \citet{Moustakas2010}. The metallicities utilize the \citet{Kobulnicky04} calibration. {\it a:} As discussed in \S~\ref{wilson}, we refer to the characteristic metallicity of NGC 4736 (9.01 $\pm$ 0.03, \citealt{Moustakas2010}) in Figure~\ref{cenmet} instead of this central value. \label{basic} 
\end{center}
\end{table*}

\subsubsection{CARMA observations} 
We observe all galaxies in the (J=1-0) transition of $^{12}$CO with CARMA in the C and D configurations. The observations presented in this paper were taken from early 2007 through March 2012. For most of the observations, the correlator configuration utilizes three dual side bands, each with 63 channels, which overlap each other for continuous velocity coverage. Six (overlapping) edge channels are removed per sideband in the data processing. The total bandwidth is 100~MHz with 1~MHz channels (5.08 \kms after Hanning smoothing). The field of view of each CARMA 19-pointing mosaic is 2.3$\am$. A few tracks for NGC 2403 and NGC 3031 were taken after the upgrade of the CARMA system and have better velocity resolution, but we smooth the data in order to be consistent with the majority of the data presented here.

\subsubsection{NRO45 observations} 
To achieve accurate total flux measurements, we also observe the galaxies using the Beam Array Receiver System (BEARS) instrument -- a multi-beam receiver with 25 beams -- on the NRO45 single dish telescope. Using on-the-fly mapping and BEARS, we sweep the galaxies with ON scans in both the RA and Dec directions, and we observe OFF (sky) pointings at positions at least 3' outside of the galaxies [as defined by the 0.12 MJy/sr contour of the 24$\micron$ emission; Koda et al., in preparation]. To reduce non-linearities in the spectral baselines, we interpolate between subsequent OFF positions and subtract the result from the intervening ON scan using the NOSTAR package developed at NRO. Each ON+OFF scan takes $\sim$1 minute, and we cover the entire galactic disks (i.e., areas much larger than the CARMA fields of view) in $\sim$40 minutes. The total bandwidth is 265~MHz, and we smooth the 500 kHz velocity resolution to 5.08 \kms~to match the CARMA data. 

\subsubsection{CO data combination} 
To measure accurate fluxes, we weight the CARMA and resampled NRO45 {\it uv}-data by their respective rms noise levels (instead of applying an ad hoc scaling factor to match the zero spacing amplitude) and image them together. This procedure is explained in detail in \citet{Koda11}. Though the single dish maps have larger spatial and velocity coverage, we use only the field of view and bandwidth of the CARMA observations. The sizes of the Gaussian beams used to image the combined {\it uv} data -- which we call the ``combined beam sizes" -- are calculated as described in \citet{Koda11} and DM12. The 1$\sigma$ rms levels and combined beam sizes are listed in Table~\ref{properties}.

Since we weight the {\it uv} data by its rms noise, we flag the NRO45 data beyond where they begin to overlap in {\it uv}-space with the CARMA data but before the single dish noise becomes prohibitively high, using a restoring combined beam size which ensures full flux recovery (relative to the single dish flux) over the velocity range of the target emission. At the largest single dish baselines, the visibilities are very noisy. From a sensitivity matching standpoint, the optimal range for the {\it uv} cutoff is in the 4-6 k$\lambda$ range \citep{Koda11}, but the actual cutoff location depends on the availability of sufficient CARMA visibilities at the smallest {\it uv}-distances. For NGC 2403, we cut at 3 k$\lambda$; for NGC 3031 and NGC 6946, 4 k$\lambda$; for NGC 4826, 5 k$\lambda$; and for NGC 4736, 7 k$\lambda$. 

Note that in DM12, we used the native resolution of the velocity channels (2.54 \kms) to optimize resolving the brightest GMCs, as opposed to detecting the full component of clouds, and we also cut NGC 6946 at 10 k$\lambda$. The resulting (slight) change in beam size and subsequent reduction in noise, in addition to the wider and more sensitive channels, account for the larger cloud sample and slightly updated derived quantities presented in this paper. The results presented for NGC 6946 here and in our previous paper are well within our quoted uncertainties in both papers.

\section{Identifying giant molecular clouds \label{randsigma} }

We use the Clumpfind algorithm \citep{Williams94} to identify giant molecular clouds in our combined CO data, as described in detail in DM12. We compare our results to those obtained with another widely used clump decomposition algorithm, \cpropse, in the Appendix. Clumpfind probes image cubes, locates coherent emission in three ($\it{x, y, v}$) dimensions by contouring the data at multiples of the rms noise, and follows the peaks down to a user-specified limiting value. We use twice the rms noise as both the increment and the minimum contour level. The algorithm output includes the clump centers, radii, and velocity widths, the latter two of which are corrected for our resolution elements as described in \S~\ref{res-section}. 

We employ a modification to Clumpfind which first searches the boundaries of existing peaks for adjacent ``inter-peak" emission pixels instead of automatically assigning such pixels to the nearest, but potentially inappropriate, peak. This avoids the situation in which a pixel is mistakenly assigned to an already-identified clump with which it has no physical contact. Imagine, for example, that the area consisting of Canada and the United States has been identified as one clump, and Mexico has been identified as a second clump. If Florida is the pixel in question, the original Clumpfind algorithm would find the Mexican peak (Pico de Orizaba) to be closest and attempt to associate the Florida pixel with the Mexican peak. The new implementation ascertains that the pixel's physical connection to the Canadian/US peak (Mt. McKinley) is more valid and avoids this misidentification. Similar treatments -- also used in conjunction with Clumpfind -- are employed in the analyses of \citet{Brunt03} and \citet{Rosolowsky05}.

\begin{figure*}
\plotone{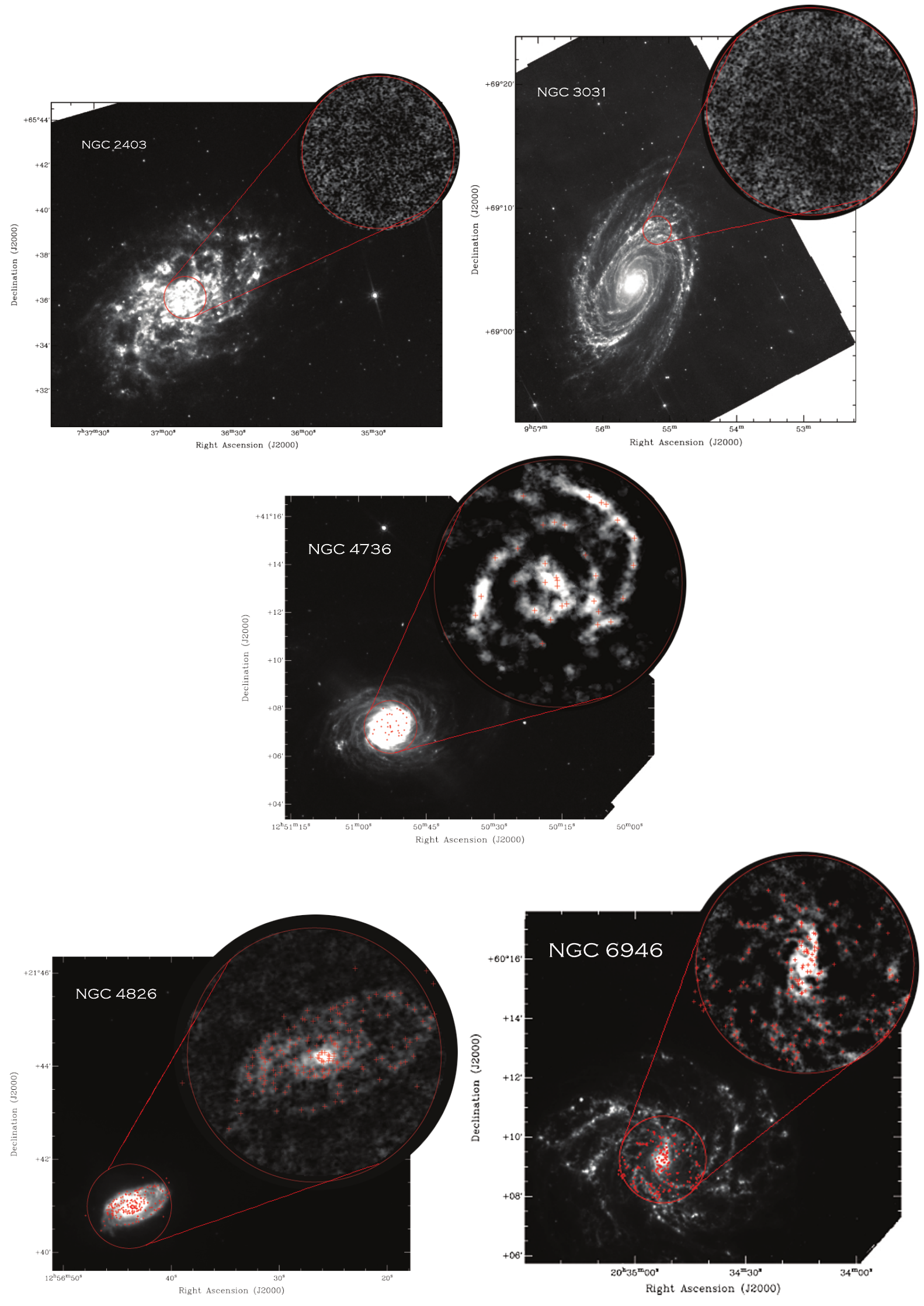}
\caption{Insets of the CO emission with the locations of all Clumpfind clouds remaining after resolution and flux cuts are shown with the 8$\micron$ image of each galaxy. \label{co-clouds} }
\end{figure*}

The output list is further pruned in a series of ways: first, we include only clumps with integrated fluxes greater than three times the integrated instrumental sensitivity over the clump volume (which changes across the field of view due to primary beam effects). Second, we eliminate clumps which are smaller than our spatial and/or velocity resolution elements. The clump samples remaining after these two criteria are met are shown in Figure~\ref{co-clouds} overlaid on 8$\micron$ images and ``zoomed in" CO integrated intensity moment zero maps of each galaxy. The CO moment maps are created by smoothing the CO cubes with an 8$\as$ Gaussian beam, clipping the smoothed emission fainter than 1$\sigma$, and using the resulting mask to create an integrated intensity image with the high resolution CO cube.

\begin{figure}
\plotone{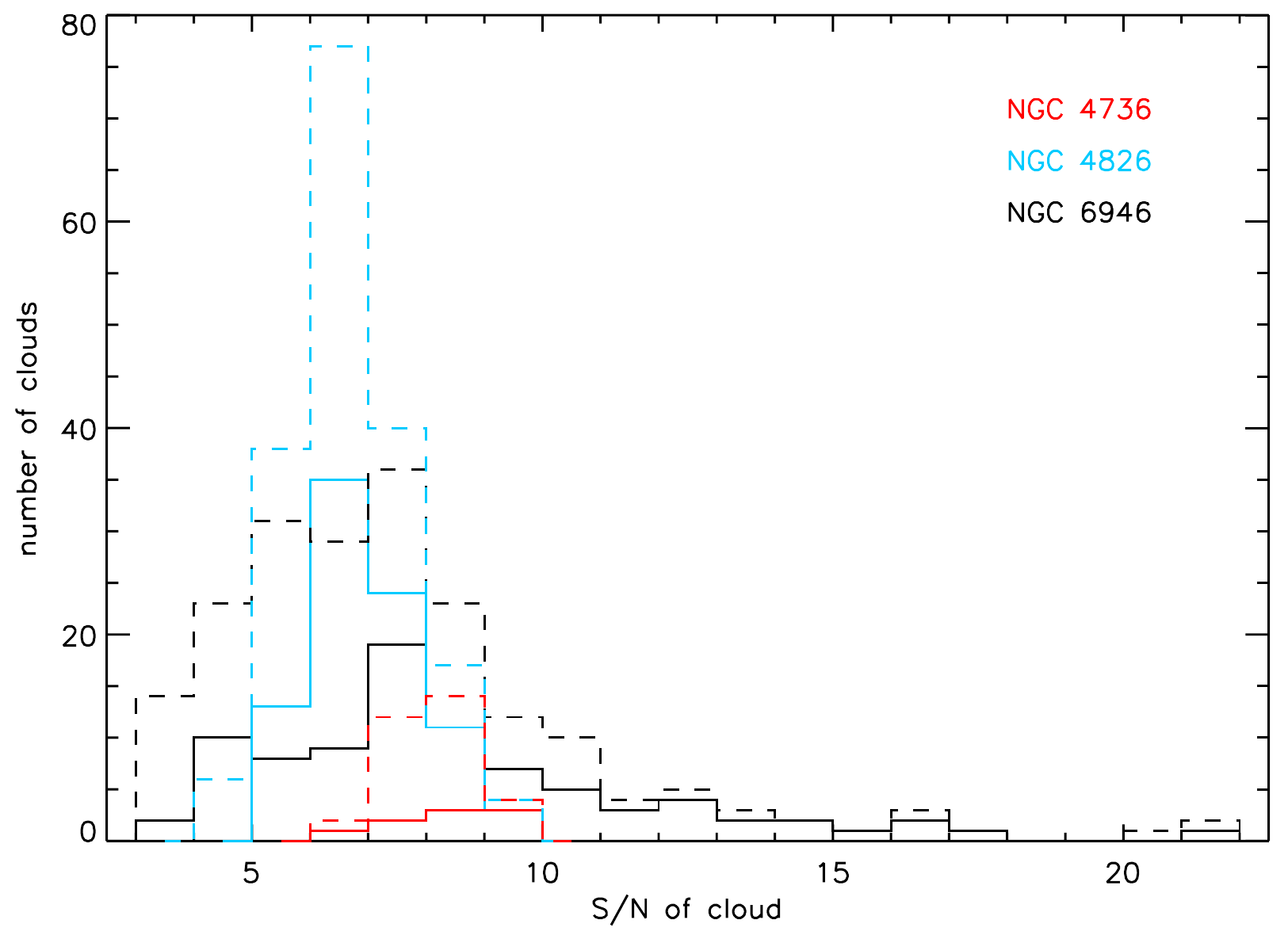}
\caption{Signal-to-noise of our detected clumps {\it (dashed lines)} and kept clouds {\it (solid lines)}. \label{sensitivity} }
\end{figure}

The percentages of the CO emission recovered in clumps after the sensitivity and resolution cuts are shown in Table~\ref{properties} for each galaxy. The emission in NGC 2403 and NGC 3031 is very faint, and even though the beam sizes ($<$30~pc) are the smallest of our sample, we detect no clumps with significance. In NGC 4826 and NGC 6946, with beam sizes of 62 and 66 pc, we recover 71\% and 65\% of the emission in clumps. In NGC 4736, this fraction goes down to 28\%; the larger beam size limits us to resolving a smaller fraction of clumps in this galaxy. As discussed in DM12, the critical density to excite CO emission is a few hundred cm$^{-3}$, and \citet{Scoville87} find that $\sim$50\% of the total H$_{2}$ mass resides in clouds less massive than a few $\times$ 10$^{5}$ \Msun~in the Galaxy (with the other half found in clouds more massive than this threshold). Thus it is likely that the emission not assigned to clumps in all five galaxies in our analysis resides in molecular clouds smaller than our resolution elements and/or fainter than our detection limits.

\begin{figure*}
\plotone{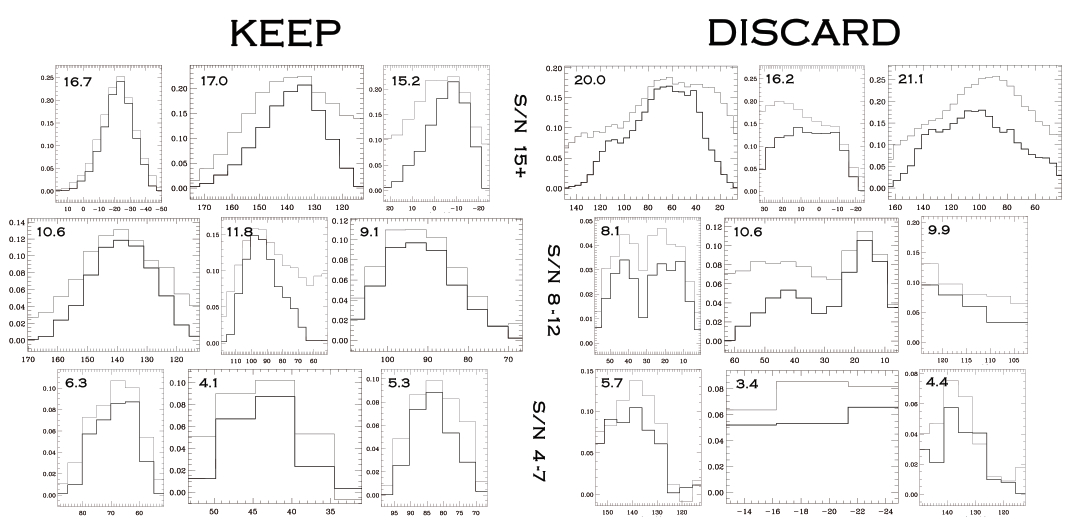}
\caption{Example velocity profiles are shown of both kept clouds {\it (left)} and discarded clumps {\it (right)} at high {\it (top row)}, intermediate {\it (middle row)}, and low {\it (bottom row)} signal-to-noise ranges. In each panel, the x-axis units are \kms, and the y-axis units are mJy beam$^{-1}$. The dark line indicates the velocity profile of the emission designated by Clumpfind to belong to a particular clump/cloud, and the gray line indicates the total emission along the same line of sight. The signal-to-noise ratio of each clump/cloud is displayed in the upper left corner of each panel. \label{examples} }
\end{figure*}

Finally, we visually inspect the velocity profile of each detected clump that passes our sensitivity and resolution cuts. If the clump profile appears to be a blend of multiple clumps (i.e., it exhibits more than one peak) or a partial clump (i.e., it does not have a regular, approximately Gaussian shape), the clump is rejected; otherwise, it is added to our final cloud sample. It is likely that we discard real GMCs -- particularly the smallest clumps with partial profiles and those in more crowded regions with blended profiles -- but we prefer to conservatively present only the clouds in each galaxy which are most likely to adhere to our assumption of virialization. We summarize the range of sensitivities of our clumps that survive the sensitivity and resolution cuts and the subset which comprise our kept clouds in Figure~\ref{sensitivity}. The faintest cloud in the final sample is detected at 3.5 times the instrumental sensitivity; 93\% of our kept clumps yield signal-to-noise ratios greater than 5. As may be expected from our conservative by-eye approach, most rejected clumps are at the low signal-to-noise end of the distribution. Examples of kept and discarded velocity profiles are shown in Figure~\ref{examples}.

Of the emission identified as being in clumps in the three brighter galaxies, we retain 40-70\% in clouds after our velocity profile assessment.
We find that the algorithm performs ``better" (that is, by our velocity profile analysis) as the beam gets smaller, with of course the exceptions to this rule being the faint galaxies NGC 2403 and NGC 3031. 

\begin{table*}[t]
\begin{center}
\caption{Observational and Derived Properties \label{properties} }
\smallskip
\begin{tabular}{cccccc}
\hline
\hline
& NGC 2403 & NGC 3031 & NGC 4736 & NGC 4826 & NGC 6946 \\
\hline
FWHM combined beam size (pc) & 28 & 29 & 78 & 62 & 66 \\
1$\sigma$ (mJy beam$^{-1}$) & 31 & 22 & 76 & 31 & 39 \\
Min./increment level (mJy beam$^{-1}$) & 70 & 50 & 150 & 60 & 80 \\
Number of clouds & 0 & 0 & 9 & 86 & 87 \\
\hline 
Average surface density (\Msun~pc$^{-2}$) & ... & ... & 120 & 140 & 170 \\
Average \xco (10$^{20}$ cm$^{-2}$ / (K km s$^{-1}$))$^{a}$ & ... & ... & 1.83 & 1.27 & 1.15   \\
Fraction of CO emission in clouds$^{b}$ & ... & ... & 12\% (28\%) & 47\% (71\%) & 38\% (65\%) \\
Total mass in kept clouds (\Msun) & ... & ... & 3.5e7 & 2.8e8 & 5.8e8 \\ 
\hline
\end{tabular} \\
$\it{a}$: Our measurement errors on the sizes and velocity dispersions of clouds translate to errors of a factor of two on the average values of \xcons. $\it{b}$: Fraction of all CO emission in clouds that survive our velocity profile analysis, i.e., clouds analyzed in this paper (fraction of all CO emission in clouds remaining after only sensitivity and resolution cuts). 
\end{center}
\end{table*}

\section{Deriving GMC Properties}

\subsection{Resolution correction \label{res-section} }

We account for the resolution elements (the spatial imaging beam and the velocity channel width) by approximating both the clouds and resolution elements as Gaussian shapes, as subtracting them in quadrature mathematically yields a Gaussian-shaped result. In reality, the channel width is a boxcar function, but we neglect this effect. In this way, we consider a cloud resolved when its deconvolved diameter is at least one beam wide (its radius is greater than or equal to half of the beam size) and its deconvolved velocity width is at least as wide as half of one (Hanning smoothed) velocity channel. Thus (from \citealt{Williams94} and DM12):

\begin{equation} 
R = \sqrt { R_{meas}^{2} -  \bigg( \frac{b}{2.355} \sqrt{2 ln \big( \frac{T}{\Delta T} \big) } \bigg) ^{2} }
\end{equation}
and \\
\begin{equation} 
\sigma_{v} =  \sqrt { \sigma_{v, meas}^{2} - \big( \frac{\Delta v_{chan}}{2} \big) ^{2} }, 
\end{equation}

\noindent where the circular radius is corrected for the Gaussian beam size ($b$) with the same peak brightness temperature ($T$) as the cloud measured at the largest projected extent of the cloud ($T = \Delta T$), $R_{meas}$ and $\sigma_{v, meas}$ are the cloud radius and velocity dispersion from Clumpfind, and $R$ and $\sigma_{v}$ are the final (resolution-corrected) radius and velocity dispersion measurements presented in this paper. In Equation 2, we follow the convention established by \citet{Williams94}, assuming a channel width that is half of one Hanning smoothed channel (or $\Delta v_{chan}$/2), though we note that alternate expressions exist (i.e., \citealt{Rosolowsky06}).

\subsection{Deriving mass and luminosity \label{lumlim-section} }

As described in detail in DM12, we derive virial masses from the cloud sizes and velocity dispersions measured using Clumpfind in a manner chosen to be consistent with the literature (e.g., S87; \citealt{Wilson90}; B08). The clouds are assumed to be virialized, and density profiles of the clouds are assumed to go as {\it r$^{-1}$}, yielding the equation typically cited:

\begin{equation}
M_{vir} = 1040~R~\sigma_{v}^{2}, 
\end{equation}

\noindent where $R$ is the radius of the cloud and $\sigma_{v}$ is its velocity dispersion. For simplicity and consistency with previous studies (particularly S87), the clouds are assumed to be spherical, as laid out in the calculations shown in DM12. \citet{Rebolledo12} address the potentially oblate/prolate nature of GMC geometries in NGC 6946 using deconvolved CO (J=1-0) and CO (J=2-1) maps and subsequently correct their GMC virial masses by factors only up to 20-30\% (i.e., less than our final uncertainties on \xcons). 
 
We calculate the CO luminosity of each cloud via 

\begin{equation} 
L_{CO} = \frac{ 13.6~\lambda_{mm}^{2}~F_{CO}~d_{pc}^{2} } { \theta_a~\theta_b }, 
\end{equation}

\noindent where $\lambda$ is the observed wavelength (in mm), F$_{CO}$ is the flux density in Jy beam$^{-1}$ \kms arcsec$^{2}$, d$_{pc}$ is the distance to the galaxy in parsecs, and $\theta_{a}$ and $\theta_{b}$ are the beam axes (in arcsec). Including a factor of 1.36 to account for helium, we calculate \xco using the equation

\begin{equation} 
X_{\text{CO}}~[\text{cm}^{-2}~\text{(K~km~s}^{-1})^{-1}] = 4.60 \times 10^{19}~\frac{ M_{vir} } { L_{CO} }, 
\end{equation}

\noindent where M$_{vir}$ is measured in solar masses and L$_{CO}$ is measured in K km s$^{-1}$ pc$^{2}$. Since the luminosity of a GMC can be expressed as its surface area multiplied by its temperature and velocity dispersion (i.e., $\pi R^{2} T \sigma_{v}$, \citealt{ScovilleSanders87}), for clouds with our smallest detected radii ($\sim$36~pc) and velocity dispersions (2.54 \kms) at 10~K (the temperature of an ``average" Galactic GMC with a similar size, \citealt{ScovilleSanders87}), our luminosity sensitivity limit is $\sim$10$^{5}$ K \kms pc$^{2}$, which is consistent with our least luminous cloud measurements in Figure~\ref{rsigma}. We propagate our errors in size and velocity dispersion through the equations for virial mass, luminosity, and \xco to obtain the uncertainties plotted in Figure~\ref{rsigma}. From these errors, we estimate that each value of \xco quoted in Table~\ref{properties} is accurate to within a (typical) factor of two, as described in more detail in the following section. Our cloud measurements are summarized in Table~\ref{params}; a sample is shown in the print edition, but the entire table is available online.

\subsection{Error estimates \label{errors-section} }

In this section, we discuss the possible sources of systematic, statistical, and intrinsic uncertainty in our measurements and the treatment of our errors. 

\subsubsection{Systematic uncertainties}

Potential systematic errors in our cloud measurements include the uncertainty inherent in apportioning emission into clouds; such systematic errors are often neglected. In our case, these errors also include any systematics inherent in our by-eye velocity profile analysis. These uncertainties are difficult to estimate and may be more likely to be important for clouds measured near our resolution limits. 

To estimate the systematics involved in relegating emission into clouds, we compare the NGC 6946 cloud sample in this paper to the CO emission obtained for the same galaxy using higher velocity resolution in DM12. The purpose of our previous analysis, with channel widths of 2.5 \kms, is to resolve the brightest clouds in the galaxy. The complementary goal of this paper is to be sensitive to more GMCs, which is achieved by using larger individual velocity channels of 5 \kms. Here we test the possibility that the clouds kept in the 5 \kms sample would be rejected if observed at 2.5 \kms resolution for having ``blended" or ``partial" profiles. We examine the three-dimensional volumes defined by each 5 \kms cloud using the emission in the 2.5 \kms cube to ask in each case if we would reject the resulting velocity profile. Of the 87 cloud volumes in the final NGC 6946 sample presented in this paper, ten high resolution velocity profiles ($\sim$ 10$\%$) would be rejected for exhibiting partial (8) or blended (2) velocity profiles. These ten clouds span the range of GMCs detected in this study; they are not only the smallest clouds. As such, their removal would not significantly change our results. 

The other potential systematic effect is our by-eye velocity profile pruning. Since ultimately we are interested in the conversion factors of individual clouds, there may be a concern that we are biasing our result by discarding emission with irregular profiles. We compare the average conversion factor derived using all clouds that survive our sensitivity and resolution cuts (briefly assuming that the virial method can be used) to that derived using only our ``kept" clouds. We find values roughly consistent with our average values, indicating that we are not systematically biasing our measurements of \xco by throwing out clouds with irregular velocity profiles. 

\subsubsection{Statistical uncertainties \label{stats} }

Our statistical errors are estimated in the manner of \citet{Wilson90}, i.e., 25\% of our spatial beam size ($\sim$15-19.5~pc) and half of our unsmoothed velocity channel width (1.3 \kms). The uncertainties on $\sigma_{v}$ for clouds measured near our velocity resolution limit are likely larger than 1.3 \kms, as the contribution of the channel shape to the observed profiles approaches that of the actual emission, but this estimate is suitable for the median of the population to our largest clouds. The \citet{Wilson90} convention for estimating uncertainties is a simple approximation and is likely secondary to the systematic uncertainties inherent in the cloud decomposition for individual clouds (particularly for our smallest GMCs). In addition to the discussion in this section, this issue is addressed more thoroughly in the Appendix.

To address the uncertainties in the measurements of our smallest clouds, we first consider the well-resolved GMCs in our sample: clouds with radii -- not diameters -- larger than the beam size (62-78~pc) and velocity dispersions greater than the width of a smoothed velocity channel (5.08 \kms). If the scatter in GMC properties is entirely due to accumulated uncertainties, as opposed to intrinsic scatter in GMC properties, the error bars should be larger than the observed scatter. For the well-resolved clouds in our sample, this is not the case; the scatter is larger than the errors on individual cloud measurements, indicating that the scatter is intrinsic to the clouds. 

Next, we can assess upper limits of the errors on the smallest clouds. As a test, if we double the estimated uncertainty in the measured cloud radii to 50\% of the beam and the velocity dispersion uncertainty to 2.6 \kms, the propagated error in \xco becomes significantly larger than the measured scatter in Figure~\ref{lcomvir}. Such large errors would be unphysical, particularly given our observation that the scatter in the well-resolved clouds is intrinsic in nature. We conclude that the uncertainties in size and velocity dispersion should be smaller than this proposed increase and consider these values to be upper limits to the appropriate uncertainties in GMC size and velocity dispersion for our smallest clouds. Note that we conservatively consider these smallest clouds to be ÒmarginallyÓ resolved even though the GMC radius (not the diameter) is smaller than the beam and the velocity dispersion (not the full velocity width) is smaller than a smoothed channel. In both cases we still actually have one to two resolution elements across each cloud. The above estimates of upper limits are therefore quite conservative, and with this caveat, we show the same errors for all clouds in Figures~\ref{rsigma} and~\ref{lcomvir}. 

\subsubsection{Intrinsic scatter}

We conclude that the intrinsic scatter of cloud properties is the dominant source of scatter in the plots. This is reasonable since, though we are conservative about our cloud selection, GMCs are not symmetric and spherical structures (i.e., \citealt{Rebolledo12}). We note that the intrinsic scatter in the GMC properties is order a factor of two, which has been observed in other local GMC studies. For clouds with measured radii larger than our beam size and velocity dispersions greater than a single smoothed channel width (i.e. very well resolved clouds), the scatter in the relationships between GMC properties (described in the following section) is larger than the uncertainties on individual clouds. 
Of course, individual cloud measurements are subject to random and systematic errors, and corresponding individual GMC uncertainties may be large. But we investigate the relationships between the cloud measurements, and these are consistent between our 2.5 \kms sample and our 5 \kms sample (as well as with the Galactic GMC sample, as described below). The scatter observed in these and other local GMC studies is also similar to the scatter that we observe here. We argue that intrinsic scatter in the GMC population drives this observed scatter and present average conversion factors for the GMCs in each galaxy with the (typical) factor of two final uncertainty.

\begin{center}
\begin{table*}
\caption{Table 3: GMC Properties \label{params} }
\medskip
\begin{tabular}{cccccccccc}
\hline
\hline
Number & RA$_{J2000}$ & Dec$_{J2000}$  & R & $\sigma_{v}$ & L$_{CO}$ (10$^{5}$) & M$_{vir}$ (10$^{5}$) & $X_{\text{CO}}$ (10$^{20}$) & Arm Class & S/N \\
 & (deg) & (deg) & (pc) & (\kms) & (K \kms pc$^{2}$) & (M$_{\odot}$) & ([cm$^{-2}$ (K \kms)$^{-1}$]) & & \\
\hline
NGC6946-1 &        308.71944 &        60.153889 & 181  & 20.1  & 340 & 761 & 1.03 & 1 & 21.9 \\
NGC6946-2 &        308.72307 &        60.155139 & 208  & 9.64  & 193 & 201 & 0.478 & 1 & 16.7 \\
NGC6946-3 &        308.71525 &        60.155000 & 140  & 20.9  & 158 & 635 & 1.84 & 1 & 17.4 \\
NGC6946-4 &        308.72027 &        60.147639 & 233  & 15.3  & 123 & 569 & 2.13 & 1 & 10.8 \\
NGC6946-5 &        308.71525 &        60.156389 & 115  & 17.7  & 87.9 & 374 & 1.96 & 1 & 14.3 \\
NGC6946-6 &        308.71972 &        60.150000 & 150  & 15.0  & 84.7 & 349 & 1.89 & 1 & 12.0 \\
\hline
\end{tabular}
\\
Properties of the GMCs resolved and analyzed in this paper; the complete table is available in the online edition. The columns include cloud positions (RA, Dec), radius, velocity dispersion, luminosity, virial mass, conversion factor (or $X_{\text{CO}}$), arm class estimate (1=spiral arm, 2=interarm region), and signal-to-noise. The uncertainties in size and velocity dispersion are assumed to be 25\% of the spatial beam size (15-19.5 pc) and half of the unsmoothed velocity channel width (1.3 \kms). As discussed in \S~\ref{errors-section}, the upper limit to the uncertainties for the smallest clouds is half of the spatial beam size (30-39 pc) and 2.6 \kms.
\end{table*}
\end{center}

\section{GMCs in Nearby Galaxies}

Our sample of resolved GMCs in nearby large spiral galaxies is the largest such sample of clouds within galaxies similar to but outside of the Milky Way. It allows us to compare these populations of clouds among substantial spiral galaxies and determine whether the Galactic conversion factor typically assumed is appropriate for all such systems. In addition, we detect GMCs within the galactic centers and within the inner disks of these galaxies, enabling comparisons of clouds in both environments with the corresponding cloud populations in the Milky Way. 

In \S~\ref{radsvels}, we discuss the properties of our resolved GMCs in the context of their counterparts in the Milky Way disk (S87) and in the disks of other Local Group dwarfs, M31, and M33 (B08). Compared to the Milky Way disk, the population of GMCs in the Galactic center may be considerably different \citep{Oka01}. As a component of our clouds is located within 400~pc of the centers of NGC 4826 and NGC 6946, we compare these clouds to Galactic center clouds in \S~\ref{section-outliers}. The GMCs in the galactic centers of our sample galaxies show only marginal deviations from the properties of the GMCs in the Galactic disk, but these are in the direction consistent with Galactic center trends as observed by \citet{Oka01}.

\subsection{GMC properties and their Galactic disk counterparts \label{radsvels} }

\begin{figure*}
\plottwo{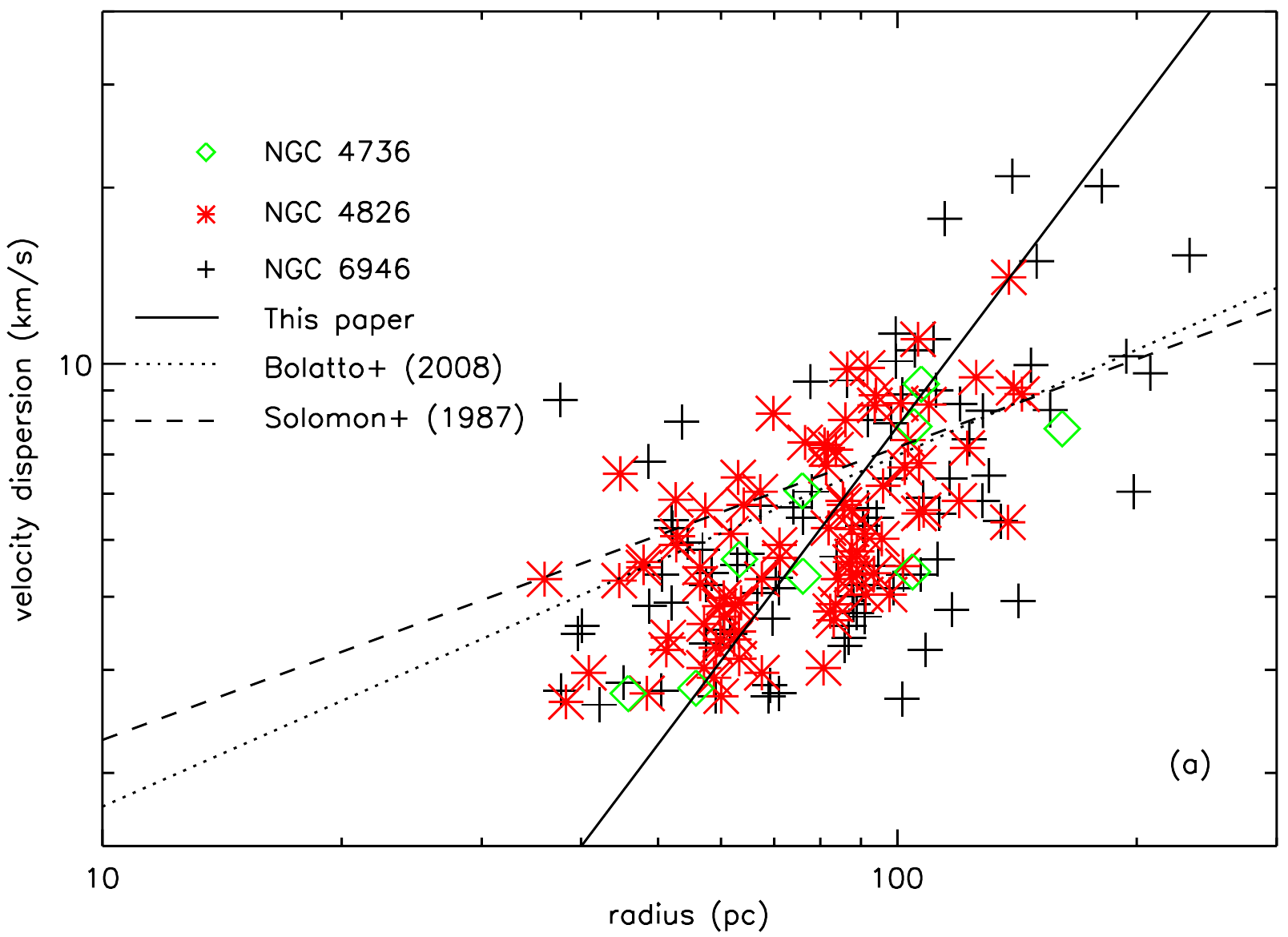}{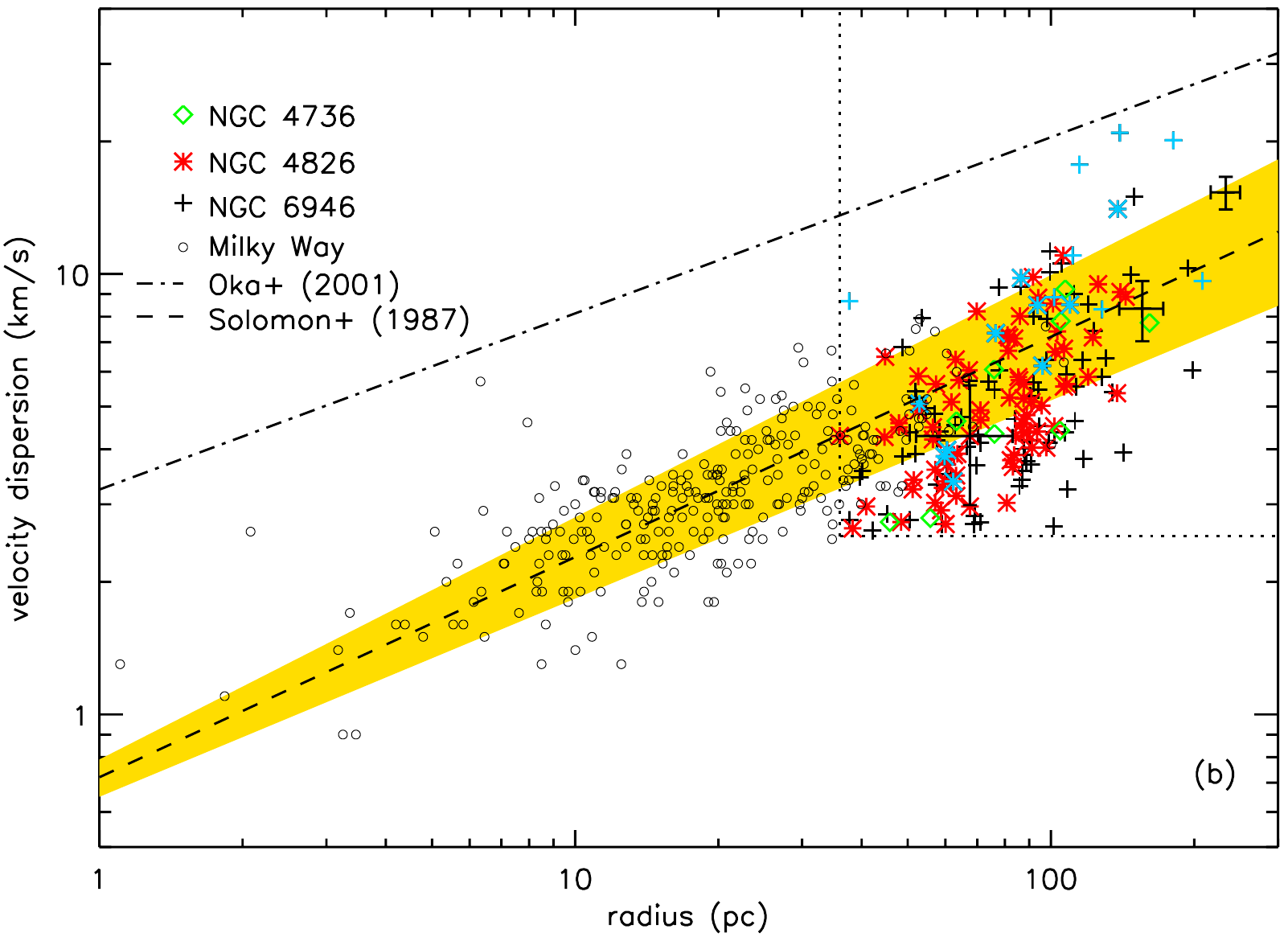}
\plottwo{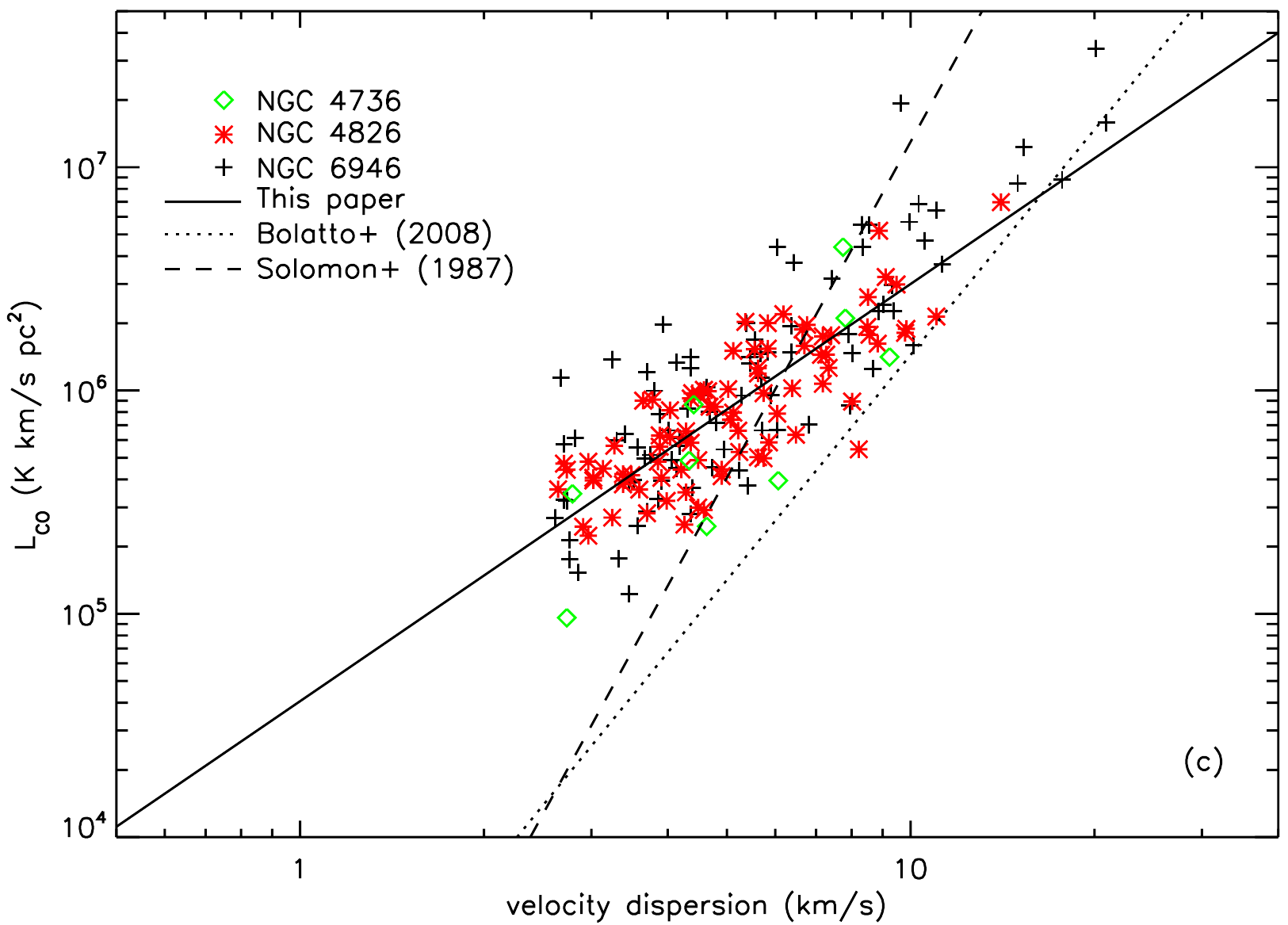}{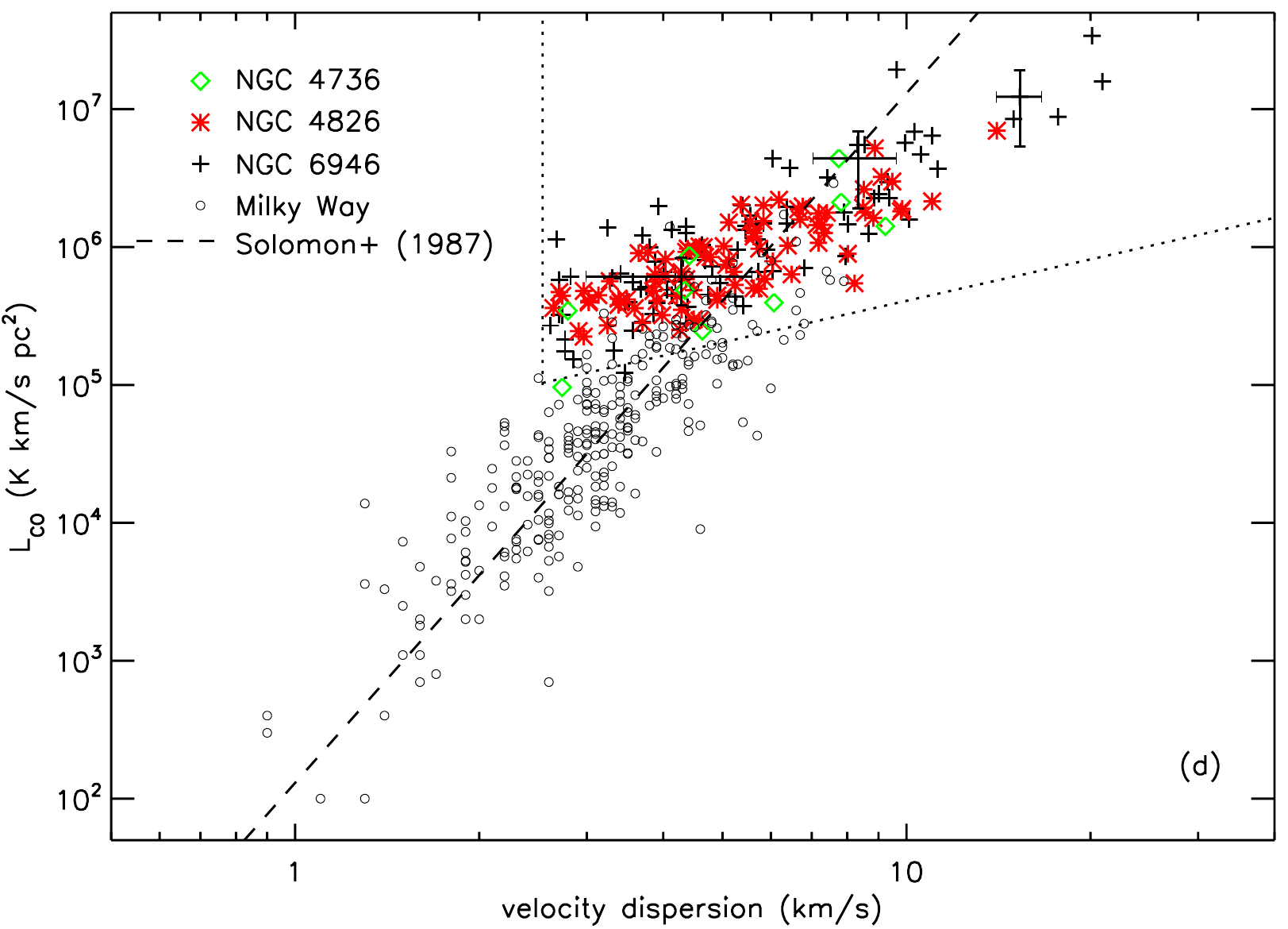}
\plottwo{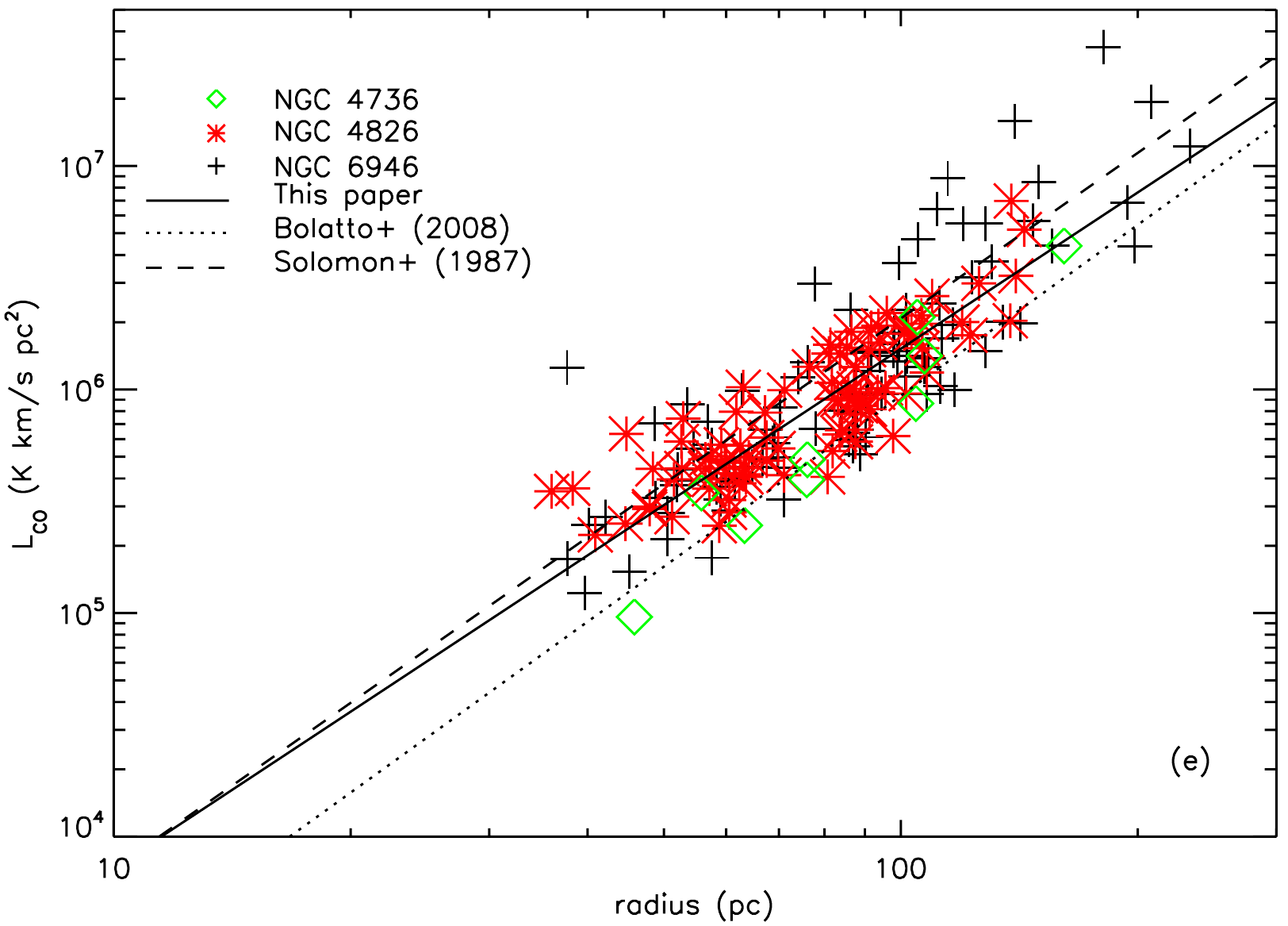}{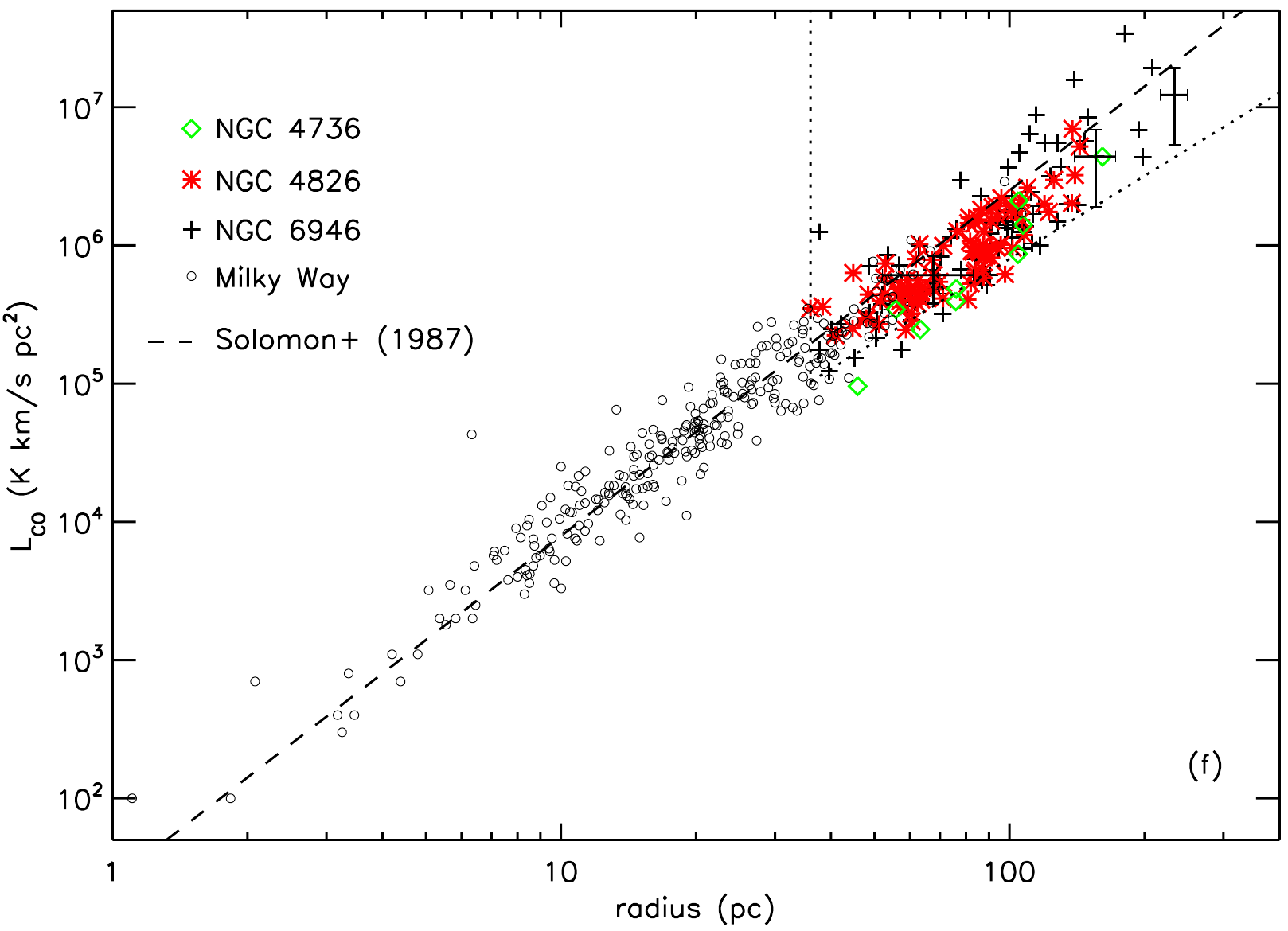}
\caption{
{\bf $\it{Top~Row:}$ } ($\it{a}$) Radius vs. velocity dispersion for our 182 detected GMCs in nearby spiral galaxies. Galaxies are coded with different symbols. The solid line shows the linear fit to the clouds in our survey galaxies (10$^{-2.71 \pm 0.49} R^{1.80 \pm 0.21}$); the fits to GMCs in the Milky Way [(0.72 $\pm$ 0.07) R$^{(0.50 \pm 0.05)}$, S87] and Local Group galaxies [(0.44 $\pm$ 0.18/0.13) R$^{0.60 \pm 0.10}$, B08] are also shown. ($\it{b}$) Our nearby galaxy GMC sample is shown with the Milky Way disk GMC sample detected by S87. The errors to the S87 fit are shown shaded in yellow. The relation shown in \citet{Oka01} from \citet{Oka98} for Milky Way GMCs at the Galactic center is added. Individual galaxy symbols are shown in blue to indicate GMCs identified within 400~pc of each galactic center. 
{\bf $\it{Middle~Row:}$ } ($\it{c}$) Velocity dispersion vs. CO luminosity for our sample of GMCs in nearby spiral galaxies. The solid line is the linear fit to all of the clouds (10$^{4.61 \pm 0.38} \sigma^{1.87 \pm 0.52}$), and the fits to GMCs in the Milky Way (130 $\sigma^{5.0}$, S87) and Local Group galaxies (645 $\sigma^{3.35}$, B08) are also shown. ($\it{d}$) The Milky Way disk GMC sample (S87) is added. 
{\bf $\it{Bottom~Row:}$} ($\it{e}$) Radius vs. CO luminosity for our sample of nearby galaxy GMCs. The solid line is again a fit to our clouds (10$^{1.54 \pm 0.88}$ R$^{2.32 \pm 0.60}$), while the relations for the Milky Way GMCs (25 R$^{2.5}$, S87) and Local Group GMCs (7.8 R$^{2.54}$, B08) are shown for reference. ($\it{f}$) The Milky Way disk GMC sample (S87) is added. In each right-hand panel, we show typical error bars for three clouds with a range of velocity dispersions as well as our sensitivity limits, which are represented by the dotted lines; luminosity limits are estimated assuming T=10~K and the minimum cloud size obtained ($R$=36~pc), as discussed in \S~\ref{lumlim-section}. 
\label{rsigma} }
\end{figure*}

The GMC radii and velocity dispersions, measured as described in the previous sections, of 182 GMCs with symbols coded by their host galaxies are plotted in Figure~\ref{rsigma}. The relations measured in previous GMC studies using resolved virial masses are shown (S87, B08). The sample is also shown relative to the Milky Way disk GMC sample (S87). From the standpoint of size and velocity dispersion, GMCs are similar among the galaxies in this study. In addition to our linear fit, which is derived from the entire cloud sample using errors in both axes, the measurements are also consistent with the overplotted S87 and B08 relationships. Our velocity and size resolution limits, as well as sizes of typical error bars, are also shown in Figure~\ref{rsigma}. We note that the linear fit to our GMCs is likely to be artificially steepened by a handful of well resolved, high $\sigma_{v}$ clouds (which will be discussed below). The fit may also be steepened by our lowest $\sigma_{v}$ clouds, which are pushing our velocity resolution and may have larger errors, but increasing the error budget for our least resolved clouds does not make a significant difference to any of the fits presented in Figure~\ref{rsigma}.

We also compare our measured GMC radii and velocity dispersions to their CO luminosities in Figure~\ref{rsigma}, and again we display the previously measured relationships for the Milky Way (S87) and for Local Group dwarfs, M31, and M33 (B08). 

Our detected GMCs generally follow the S87 Galactic disk relationship between velocity dispersion and CO luminosity within the range in $\sigma_{v}$ where we overlap with S87. We note that though the relationship shown by S87 is steeply dependent upon the velocity dispersion (L $\propto$ $\sigma^{5.0}$), the Galactic points are also consistent with the shallower slope measured by B08 for in nearby dwarf galaxy GMCs (L $\propto$ $\sigma^{3.35}$), the extrapolation of which also describes well the largest and best resolved clouds in our sample ($\sigma_{v}$ $>$ 10 \kms). Again, at the low $\sigma_{v}$ end, our measurements are pushing our velocity resolution limit, which may serve to artificially flatten the trend observed in the middle panels of Figure~\ref{rsigma} since the measured CO luminosities also depend on $\sigma_{v}$. We also note that the scatter at low $\sigma_{v}$ is consistent with the scatter observed in the S87 disk GMC clouds around the S87 relation. The overall trend between velocity dispersion and luminosity is in fact very similar to that seen by \citet{Rosolowsky05} in NGC 4826 for a sample of 25 clouds detected in $^{13}$CO. 

In the bottom panels of Figure~\ref{rsigma}, extragalactic GMCs are shown to exhibit sizes and CO luminosities which are very consistent with Milky Way GMCs. They are generally more luminous than the trends defined by the (mostly dwarf) Local Group sample for a given size, though this may be an effect of our sensitivity limit. The scatter in this relation is the smallest -- and the most similar to the Milky Way scatter -- of the trends shown in Figure~\ref{rsigma}. 
  
\subsection{GMCs with elevated velocity dispersions \label{section-outliers} }

For the GMCs detected within 400~pc of the centers of NGC 4826 and NGC 6946, GMCs in the Galactic center may provide a better counterpoint for discussion. In general, our GMCs are consistent with the Galactic disk clouds, but a handful of GMCs exist which deviate slightly from the trends followed by most Galactic disk clouds; these may be possible counterparts of the Galactic center GMCs. Here we compare our GMCs with the Galactic center GMCs as measured by \citet{Oka01}.

Using two samples of Galactic center clouds, \citet{Oka01} detect GMCs with systematically larger velocity dispersions for a given size compared to the S87 GMC sample in the Galactic disk (Figure 2b). They conclude that Galactic center GMCs are in pressure equilibrium with their environment and are close to gravitational instability, a condition shown to lead to clouds which exhibit the same scaling relationships as gravitationally bound clouds \citep{Chieze87}. \citet{Oka01} also suggest that the velocity dispersions of Galactic center clouds anticorrelate with their gravitational stability, such that GMCs which have dissipated much of their random motion and are left with the smallest velocity dispersions are the most gravitationally unstable ones thus most likely to collapse and initiate star formation. (We address the gravitational stability of our GMCs in \S~\ref{section-alpha}.)

The relationship measured by \citet{Oka01} is shown in panel (b) of Figure~\ref{rsigma}. They find a consistent trend (and dispersion around the trend) using two samples of clouds with different resolution; our instrumental resolution is consistent with their 1.2m sample clouds, which exhibit radii of $\sim$60-170~pc and are measured with a similar velocity resolution (2~\kms) to the one we use in this paper. Our detected GMCs within the central 800~pc of each galaxy (i.e., those having centers less than 400~pc from the galactic center) are highlighted in blue. A few central clouds in NGC 6946 and one central cloud in NGC 4826 are displaced upward from the radius-$\sigma_{v}$ trends defined by the rest of the clouds in each disk by amounts significant compared to the error in $\sigma_{v}$. The remainder of the detected central GMCs, especially in the more flocculent galaxy NGC 4826, have velocity dispersions indistinguishable from the rest of the inner disk clouds in that galaxy. This result contradicts the finding of \citet{Rosolowsky05}, though the spatial resolution in that study is twice the resolution presented here. We are likely resolving blended clouds -- with $\sigma_{v}$ of 10-50 \kms -- into smaller structures. As discussed in DM12, NGC 6946 is known to host a molecular bar \citep{Ball85}, but only one cloud with an elevated velocity dispersion ($\sigma_{v} > 12$ \kms) is coincident with the molecular bar, so the bar is likely not responsible for these few increased velocity dispersion clouds. Thus, we do not find a significant population of clouds with elevated $\sigma_{v}$ for a given size as seen by \citet{Oka01} in the galactic centers of nearby spiral galaxies; the majority of extragalactic GMCs instead are consistent with the relation derived from the Galactic disk. 

At high $\sigma_{v}$ (i.e., for clouds with velocity dispersions greater than $\sim$10 \kms), we also find that these few central clouds -- particularly the handful at the center of NGC 6946 -- deviate from the S87 velocity dispersion-luminosity trend. These GMCs are underluminous for a given velocity dispersion compared to the extrapolation of the Galactic sample, bringing them precisely in line with the extrapolation of the B08 Local Group relationship at large $\sigma_{v}$. This result also supports the finding of the same for high-$\sigma_{v}$ clouds by \citet{Rosolowsky05} in the center of NGC 4826. 

\section{\xco and Virial Parameters}
\subsection{Virial masses and \xco}

\begin{figure*}
\plotone{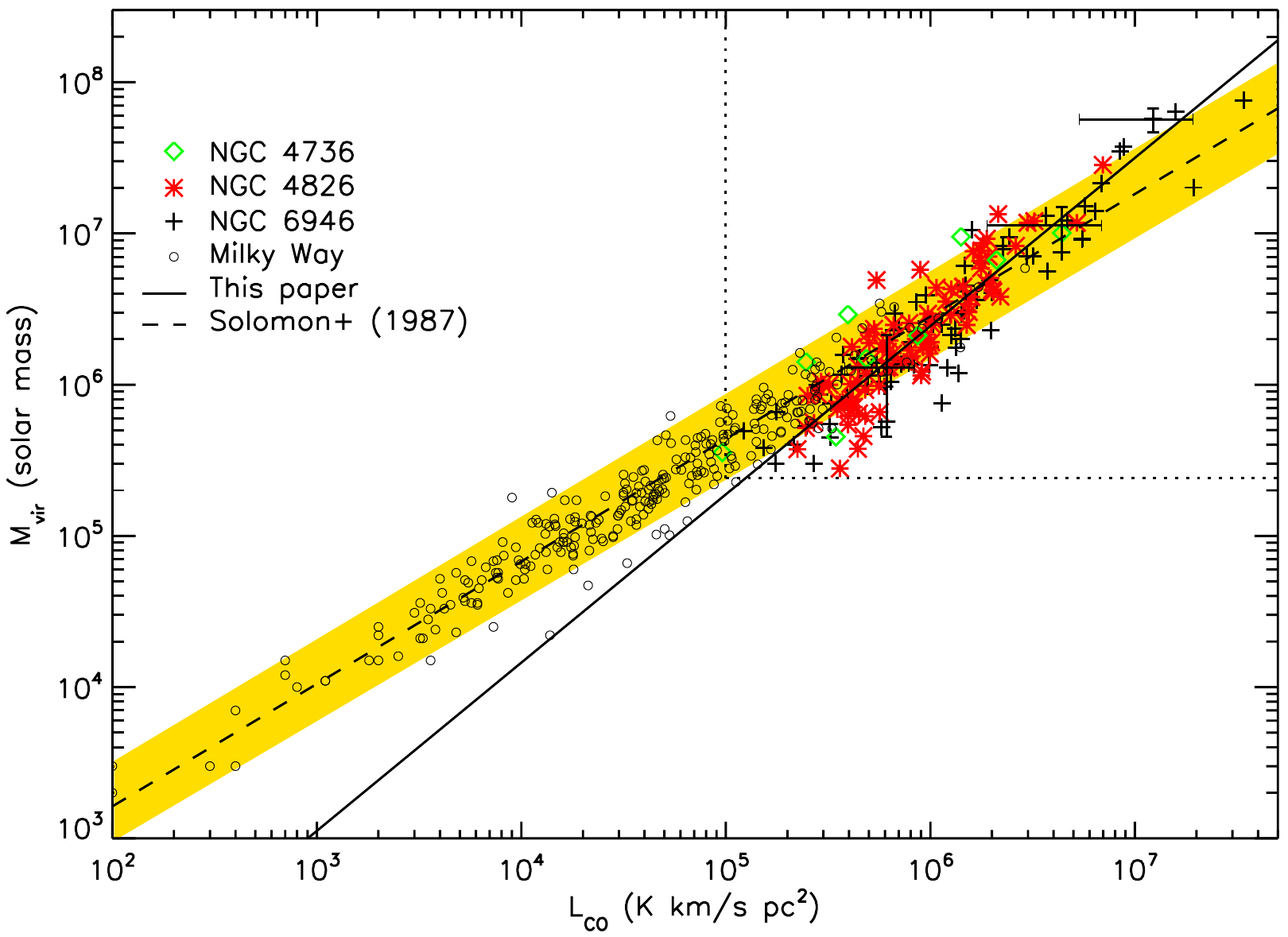}
\caption{CO luminosity vs. virial mass for our 182 detected GMCs in three nearby spiral galaxies. The solid line is the error-weighted fit to the clouds in all three galaxies (10$^{-0.29 \pm 1.06} L_{CO}^{1.11 \pm 0.29}$), and the fit to GMCs in the Milky Way (39 L$_{CO}^{0.81}$, S87) 
is also shown. Our sensitivity limits are shown by the dotted lines (assuming T=10~K for the luminosity limit, as discussed in \S~\ref{lumlim-section}), and typical error bars for three clouds with a range of velocity dispersions are also plotted. The yellow shading indicates a factor of two range in the S87 value of \xcons, which encompasses most of our clouds. \label{lcomvir} }
\end{figure*}

\begin{figure*}
\plotone{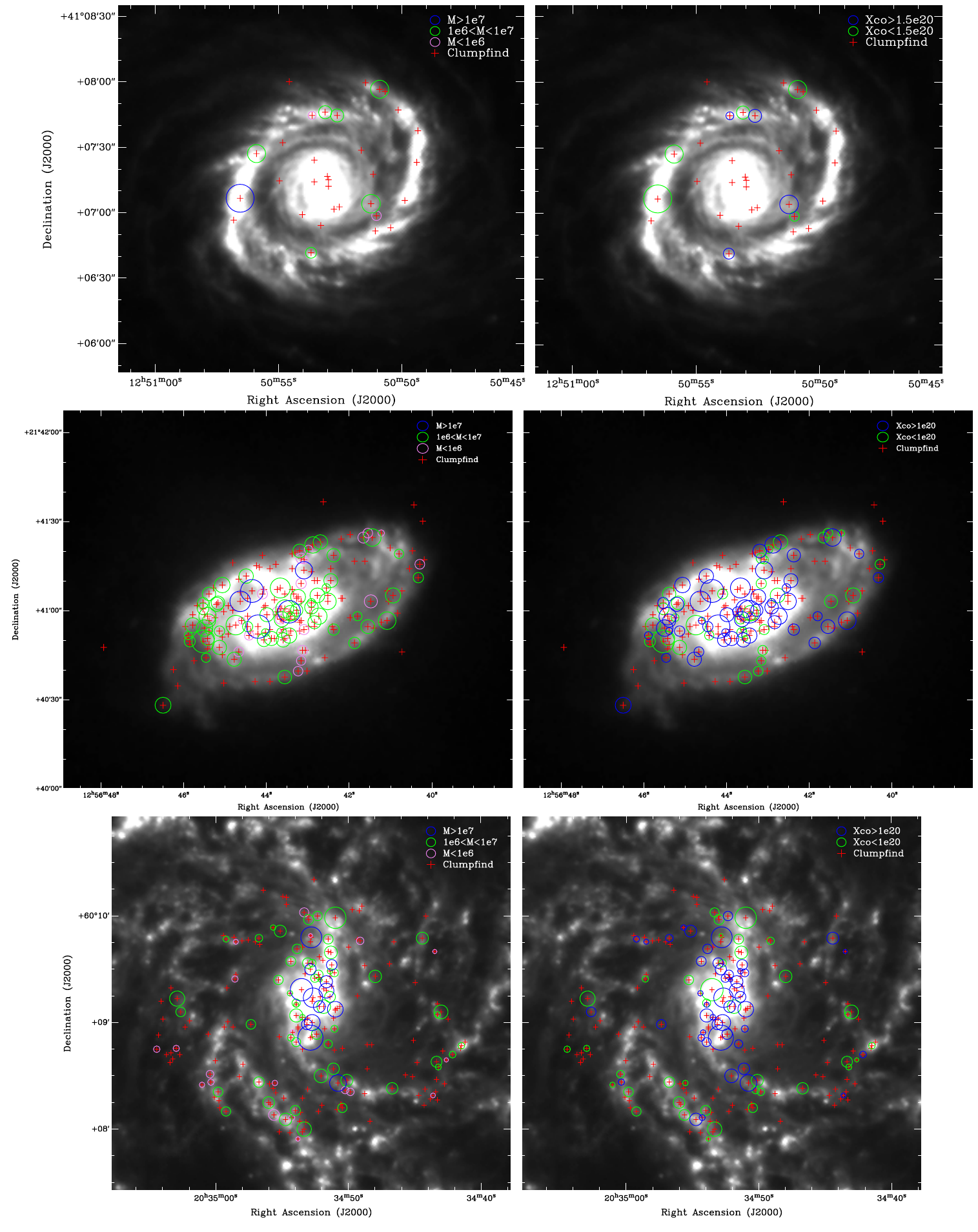}
\caption{Maps of radius, mass, and \xco distribution in (from top to bottom) NGC 4736, NGC 4826, and NGC 6946 overlaid on 8$\mu$m images. All resolved Clumpfind detections that pass our sensitivity and resolution cuts are shown with red crosses, and those which exhibit regular velocity profiles (and, as such, are included in our analysis) are also indicated with circles. \label{maps1} }
\end{figure*}

The goal of this study is to determine the virial mass based conversion factor, which if constant for each galaxy (or between galaxies) will be so regardless of the masses and luminosities of the individual clouds. The luminosities and virial masses of the individual resolved GMCs in our sample galaxies are shown in Figure~\ref{lcomvir}. The average values of the conversion factor within each galaxy, listed in Table~\ref{properties}, are all consistent within a factor of two with the Galactic value of 2 \xcounits, and average cloud mass surface densities (also listed in Table~\ref{properties}) range from 120 to 170 \Msun~pc$^{-2}$; the value measured by S87 is 170 \Msun~pc$^{-2}$ for the clouds in the Milky Way disk. As the conversion factor goes as $\sigma_{v}/R$ (assuming constant temperature, see \S~\ref{section-alpha}) and the surface density goes as $\sigma_{v}^{2}/R$, we observe some correlation between the two quantities within individual clouds; a similar relationship has been observed by \citealt{Heyer09}. Maps indicating the distribution of cloud masses, sizes, and conversion factors are shown in Figure~\ref{maps1}. 

When comparing L$_{CO}$ and M$_{vir}$, we find remarkable agreement between the GMCs probed in this paper and Galactic GMCs with masses above a few $\times$ 10$^{5}$ \Msun, below which the clouds studied in B08 resemble the Galactic GMCs. The scatter within each individual galaxy is also similar to the scatter in the S87 points in Figure~\ref{lcomvir}. A fit to our clouds yields a similar slope to that found in B08, consistent with their picture that GMC virial masses and CO luminosities are linearly related.
In other words, we find that GMCs in nearby large spiral galaxies appear to exhibit constant conversion factors as a function of cloud mass, consistent with those in Local Group spirals and dwarfs. The average conversion factor in each galaxy is within a factor of two of the average Milky Way \xcons. 

We note that the galaxies for which we detect central clouds (i.e., at galactic radius less than 400~pc) -- NGC 4826 and NGC 6946 -- exhibit lower average conversion factors than NGC 4736, where we only detect clouds at larger radius (more than 400~pc), though the cloud sample in this latter galaxy is much smaller than for the galaxies with central clouds.

\subsection{Virial parameter \label{section-alpha} }

A typical way to address the degree of gravitational ``boundness" within individual GMCs is to define a virial parameter, $\alpha_{vir} = M_{vir}/M_{lum}$ \citep{Bertoldi92, Hirota11, Rebolledo12}. Various studies have suggested that the degree of gravitational binding is set by internal cloud velocity dispersions \citep{Oka01, Hirota11}, and those clouds with suppressed velocity dispersions are able to collapse (and form stars). Clouds are more unstable to gravitational collapse as $\alpha_{vir}$ decreases. 

\begin{figure}
\plotone{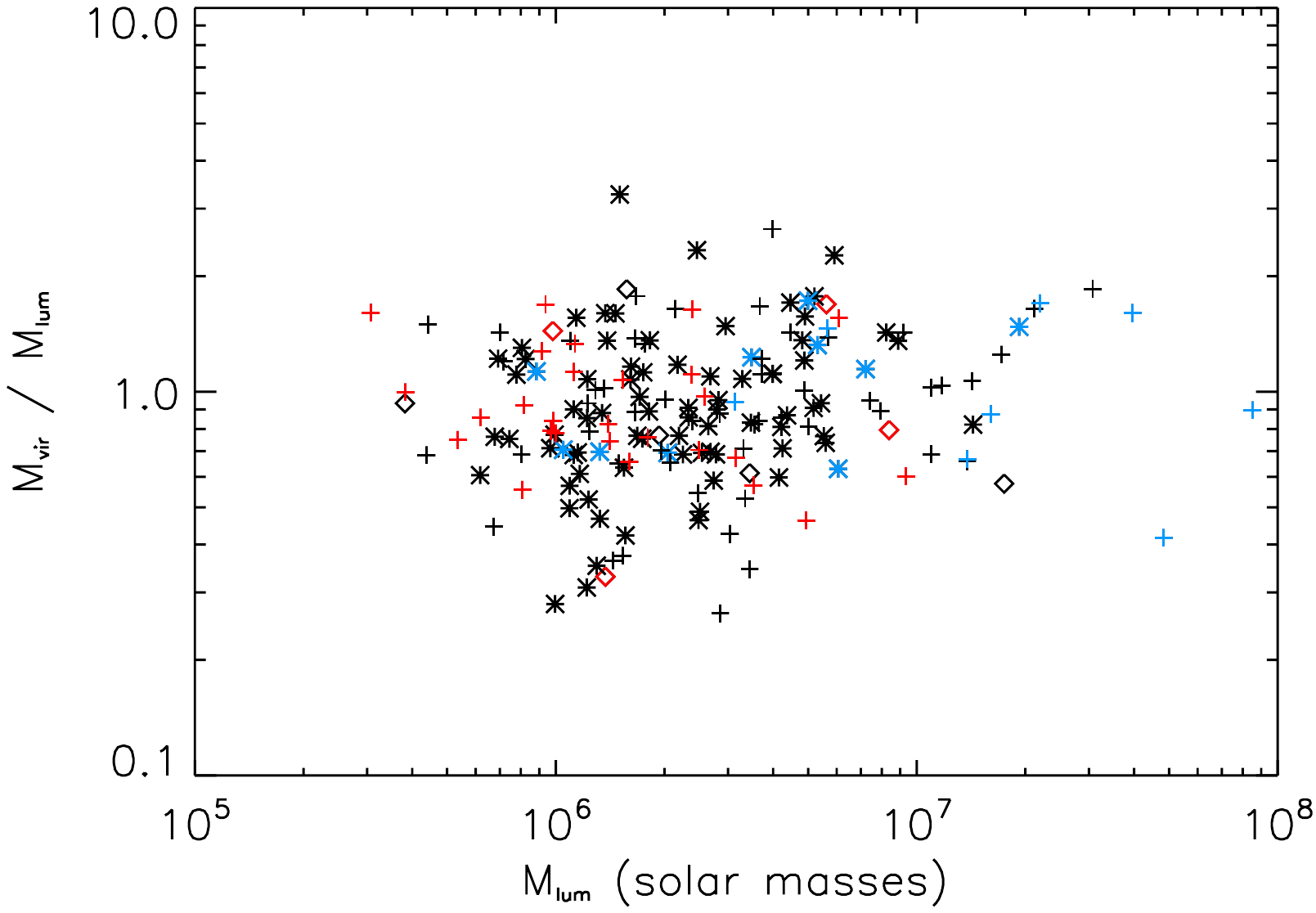}
\plotone{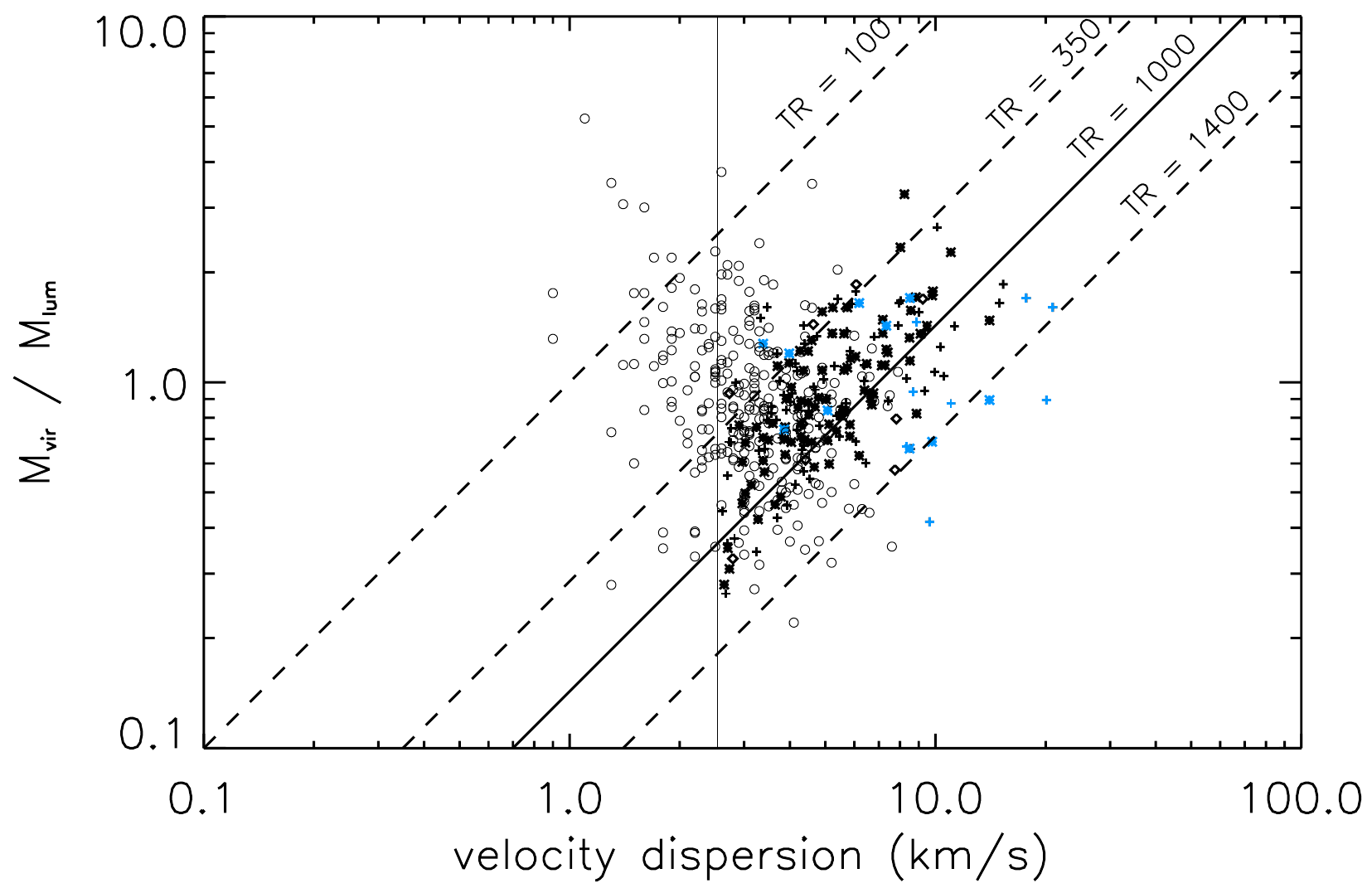}
\caption{Virial parameter correlated with luminous GMC masses and cloud velocity dispersions. In the top panel, blue points indicate GMCs in central regions (within 400~pc of each galactic center), black symbols indicate clouds located on spiral arms, and red points are clouds in interarm regions (see the text in \S~\ref{wilson} for a description of how cloud locations are established). The specific symbols are the same as in Figure 2. In the lower panel, blue points are again shown to indicate the central clouds, and the specific symbols are repeated from Figure 2. The small open circles indicate Milky Way clouds (S87). \label{alpha} }
\end{figure}

We estimate individual GMC luminosity-based masses using the average values of \xco listed in Table 2. \xco is calibrated to yield $\alpha_{vir}\sim$~1, so only relative trends are informative; for instance, in the upper panel of Figure~\ref{alpha}, no correlation is observed between the virial parameter and the luminosity-based masses. However, our average conversion factors are all within a factor of two of the Galactic value of \xco (as derived from non-virial mass arguments), and the relationship that we find between cloud mass and luminosity is linear, supporting our assumption that our GMCs are virialized (and thus most likely $\alpha_{vir}\sim$~1). 

The colors in Figure~\ref{alpha} indicate an estimate of the GMC environments (described further in \S~\ref{wilson}), thus separating clouds within 400~pc from the galactic centers (blue), interarm clouds (red), and clouds on spiral arms (black); no correlations in $\alpha_{vir}$ are seen with environment. Thus, not only are our clouds most likely bound, this ``boundness" does not appear to correlate with the spiral arm/interarm environment. 

In the lower panel of Figure~\ref{alpha}, we plot the velocity dispersion vs. $\alpha_{vir}$; our limited velocity resolution (represented by the vertical line at $\sigma_{v}$=2.54 \kms) produces an apparent positive correlation between the two quantities. Such a correlation is also seen in other recent studies \citep{Hirota11, Rebolledo12}. This is likely an artifact due to the resolution limits. As pointed out by \citet{Rebolledo12}, $\alpha_{vir}$ goes as the velocity dispersion such that

\begin{equation} 
\alpha_{vir} \propto \frac{M_{vir}}{M_{lum}} \propto \frac{R \sigma_{v}^{2}}{T \sigma_{v} R^{2}} \propto \frac{\sigma_{v}}{TR},  
\end{equation}

\noindent where, as in Eq. 1, $T$ is the peak brightness temperature. The essence of Equation 6 has also been predicted by, for instance, S87 and \citet{ScovilleSanders87}, who show that the conversion factor ($\alpha_{vir}$ in different units) should go as $\rho^{\frac{1}{2}} / T$ [since $\rho \propto M_{vir}/R^{3} \propto R \sigma_{v}^{2}/R^{3} \propto \sigma_{v}^{2}/R^{2}$, or in other words, $\alpha_{vir} \propto \sigma_{v}/TR$]. $R$ and $\sigma_{v}$ are independently observable quantities and individually produce artificial cutoffs in these plots due to their resolution limits, even though those two parameters have an intrinsic correlation.

We calculate virial parameters for the S87 Milky Way disk clouds (assuming the average value of \xco in that study, as calculated with our Equation 5, 2.6 \xcounits) and show them in Figure~\ref{alpha} as well. Due to our resolution limits, the S87 clouds are generally smaller than the GMCs presented here, so the population is seen to shift to systematically lower $\sigma_{v}$ in Figure~\ref{alpha}. 
We see no positive correlation [nor do we see a negative correlation, as suggested by \citet{Oka01}] with the inclusion of the S87 data, which are sensitive to velocity dispersions down to 1 \kms. To guide the eye, we include lines of constant $TR$, calculated from Equation 6, in the lower panel of Figure~\ref{alpha}. 

Most of the GMCs presented in this study fall between $TR$=350 and $TR$=1000, or GMCs with radii of $R$=35~pc to $R$=100~pc, assuming constant $T$=10~K temperatures. The blue points again indicate GMCs within 400~pc of each galactic center; their distribution is well mixed with that of the inner disk clouds with the exception of the central NGC 6946 clouds, which tend to exhibit higher velocity dispersions for a given value of $\alpha_{vir}$. The relative distribution of virial parameters at the galactic centers indicates that GMCs exhibit similar gravitational stability in these regions compared to GMCs in the inner disks. Varying physical conditions, producing a range of values of $\sigma_{v}$ and $TR$, can still return the same (narrow) range of virial parameters, as indicated in Figure~\ref{alpha}. 

\section{Metallicity}
\label{wilson}

We address the potential metallicity dependence of the conversion factor at the high metallicity end of such relations in two ways: first, we compare the central, or ``nuclear", metallicities of the galaxies (measured by \citealt{Moustakas2010}) to our measured average conversion factors. Second, we can address the metallicity question within individual galaxies and investigate whether metallicity-based predictions of the conversion factor agree with our observations in a more resolved (radial) sense. The range of central galaxy metallicities presented in this paper is narrow and their gradients (as measured by \citealt{Moustakas2010}) are fairly shallow, so any correlations between metallicity and the conversion factor are unlikely to have much dynamic range. Yet, as most previous studies deriving the conversion factor by resolving individual GMCs have been limited to nearby, more metal-poor systems (i.e., \citealt{Wilson95, Arimoto96, Israel97, Boselli02-xco-mets, Leroy11}), our observations yield a sense of the scatter at the opposite end of the metallicity-\xco relation. 

\begin{figure}
\plotone{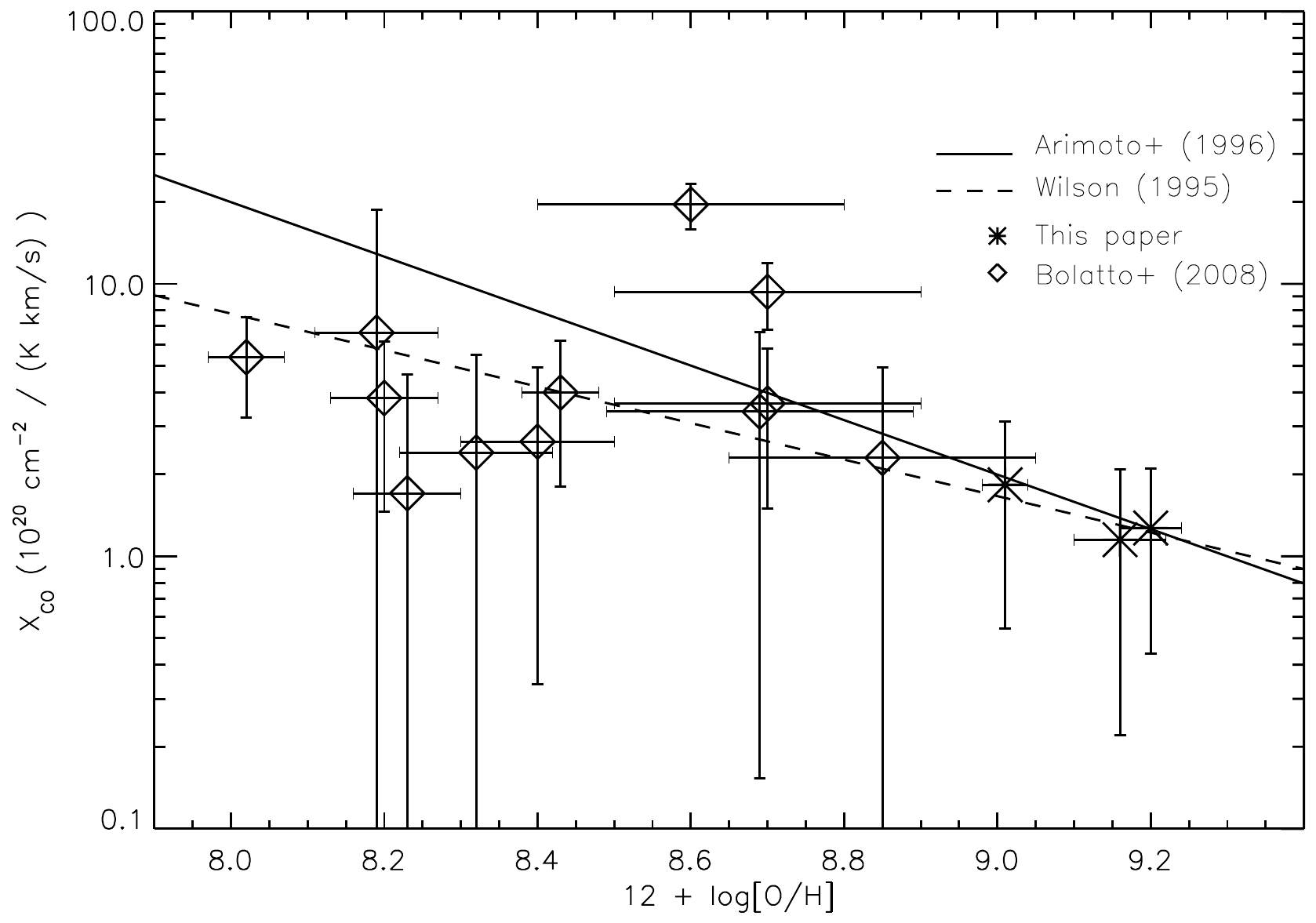}
\caption{The average conversion factors measured in the galaxies in this study are shown as a function of the metallicity of each galaxy from \citet{Moustakas2010} (see the discussion in the text for the metallicity definitions). Generally lower metallicity galaxies from B08, as well as the metallicity-\xco relations of \citet{Wilson95} and \citet{Arimoto96}, are shown for context. We show the relation of \citet{Wilson95} scaled for a Galactic conversion factor of 2 $\times 10^{20}$. \label{cenmet} }
\end{figure}

In Figure~\ref{cenmet}, we show the average conversion factor as a function of the galaxy metallicity \citep{Moustakas2010}. We show the central metallicities for NGC 4826 and NGC 6946, for which we detect central GMCs, and the characteristic metallicity for NGC 4736, where we detect only non-central GMCs. We show the values derived with the theoretical strong line calibration of \citet{Kobulnicky04} to be consistent with the metallicities derived from bright oxygen lines used in B08. We also show the lower-metallicity systems resolved by B08 and two metallicity-\xco relations from the literature \citep{Wilson95, Arimoto96} derived using virial measurements and calibrated using oxygen abundances in HII regions in lower metallicity galaxies and M31 (as well as the Milky Way and M51 in \citealt{Arimoto96}). The predictions describe well the average conversion factors and metallicities of our galaxies (as well as those of B08) and are also consistent with each other at the high metallicities of our galaxies. The metallicity values themselves, though, are a large source of uncertainty; changing the calibration to the \citet{Pilyugin05} empirical calibration quoted in \citet{Moustakas2010} results in a difference of $\sim$0.6 dex in the metallicities shown. 

\begin{figure}
\plotone{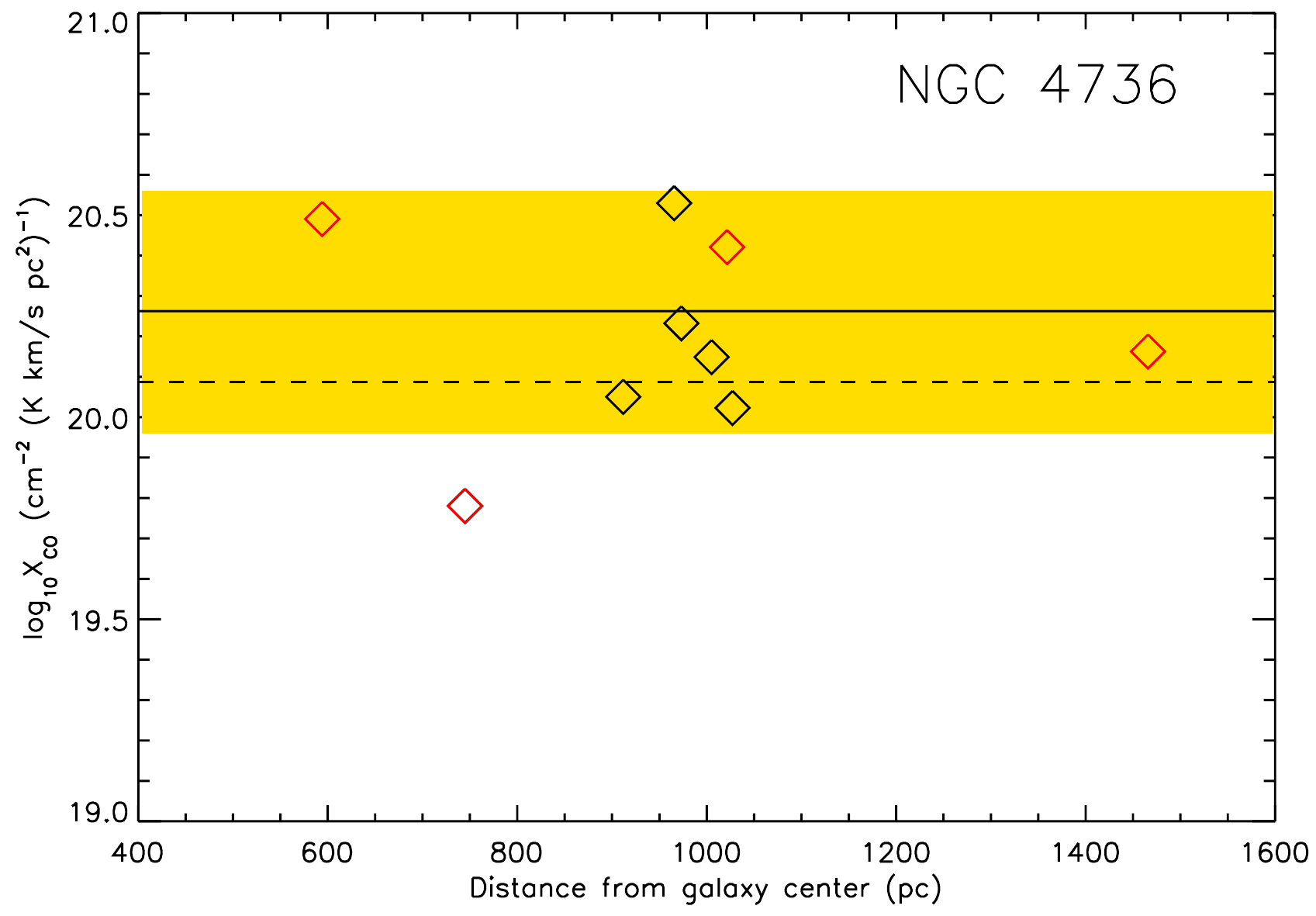}
\plotone{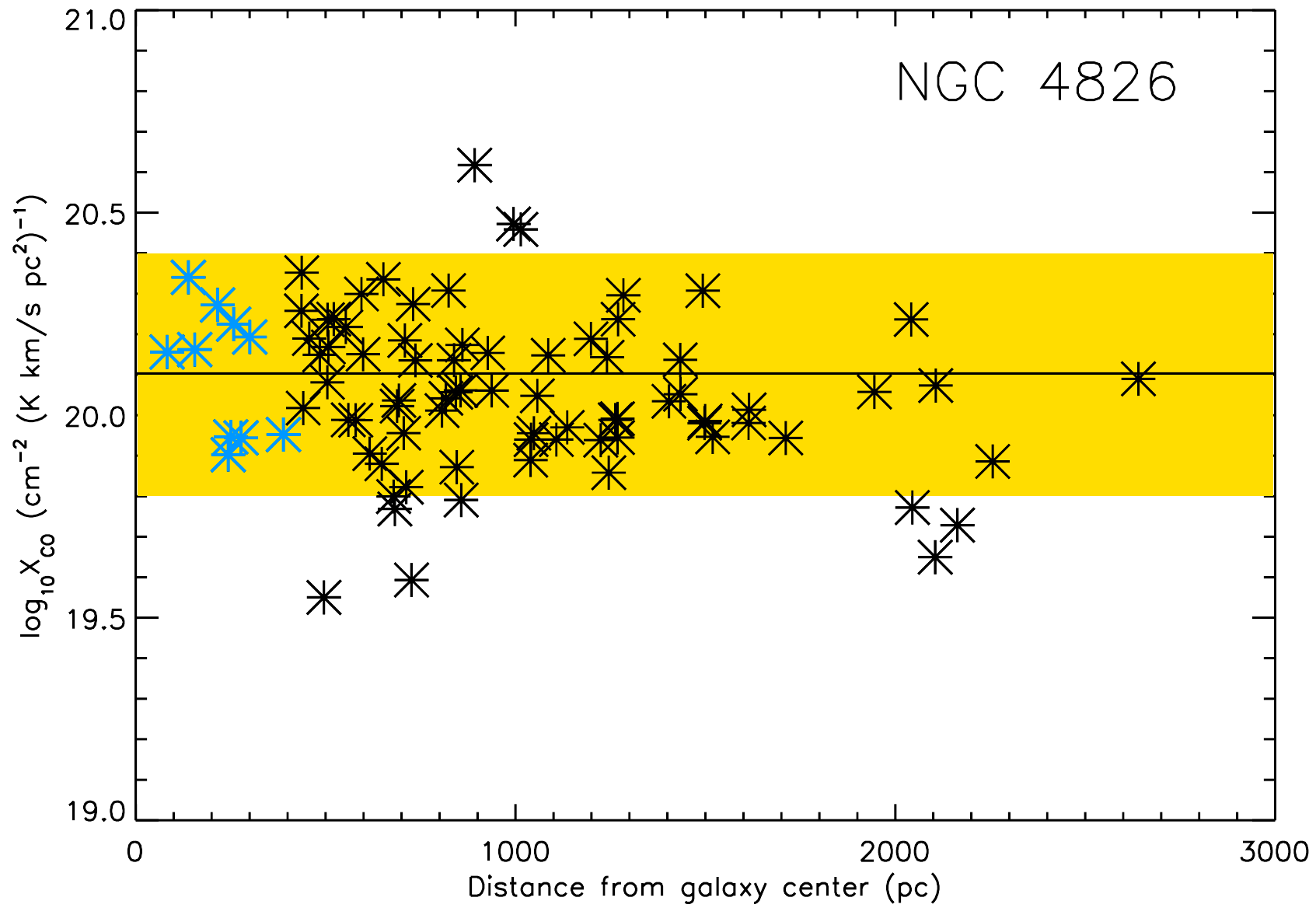}
\plotone{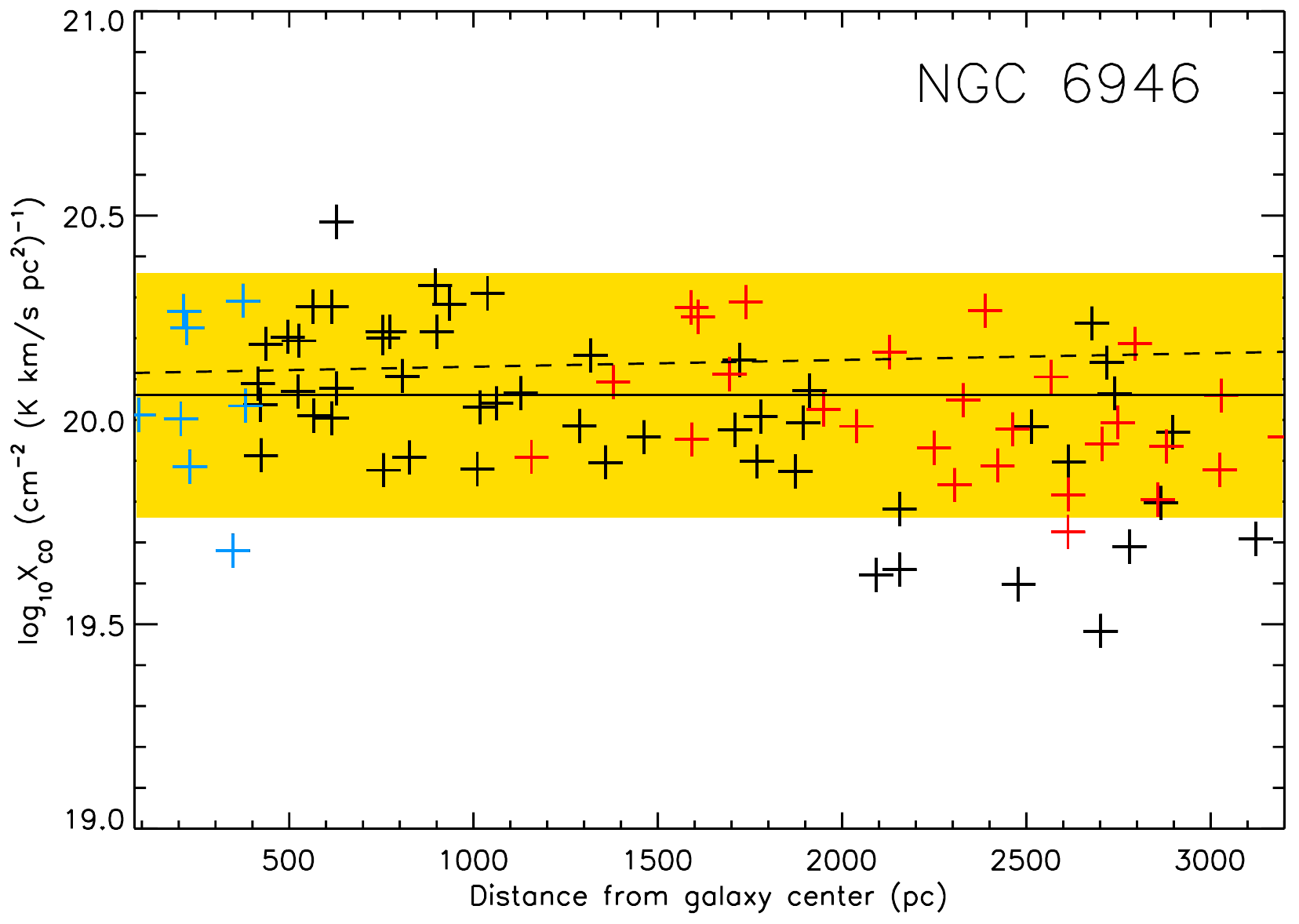}
\caption{Radial profiles of \xco for each of the galaxies in this study. In NGC 4736 and NGC 6946, GMCs are separated into arm (black symbols) and interarm (red symbols) clouds. Blue points are central points, as in Figures~\ref{rsigma} and~\ref{alpha}. Predictions for the conversion factor from the metallicity-\xco relation of \citet{Wilson95} ($\it{dashed~lines}$), derived using the central metallicities and radial gradients from \citet{Moustakas2010} with the \citet{Kobulnicky04} values {\bf (where they are available)}, are also shown. Galactocentric distances are corrected for the inclination angles published in \citet{Moustakas2010}. Average (constant) values of the conversion factor in each galaxy are indicated by the solid lines, and the yellow shading in each panel indicates a factor of two change in \xcons. \label{radialxco} }
\end{figure}

In Figure~\ref{radialxco}, we show radial ``profiles" of the conversion factor measured in individual clouds, our average (constant) conversion factor values, as well as the predictions for the conversion factor from \citet{Wilson95} (scaled for a Galactic conversion factor of 2 $\times$ 10$^{20}$) assuming the central metallicities and gradients published by \citet{Moustakas2010}. The shading indicates a range in \xco of a factor of two from our derived values. For NGC 4736 and NGC 6946, we also identify the brightest regions of the spiral arms using Spitzer IRAC 8$\mu$m imaging (i.e., Figure~\ref{maps1}); by comparing these regions to our GMC central positions, we can estimate roughly which clouds are located on spiral arms and which are located in interarm regions. We include our arm classifications in Table~\ref{params} for reference. NGC 4826 is a flocculent galaxy, so we make no assumptions about the arm and interarm regions and simply present \xco as a function of radius. Additionally, no metallicity gradient is published in \citet{Moustakas2010} for this galaxy, and so no corresponding radial prediction for the conversion factor is made. Galactocentric distances have been corrected with the inclination angles published in \citet{Moustakas2010}. 

Recent studies using dust maps derived from mid- and far-infrared emission and CO (2-1) maps suggest that a significant drop in the conversion factor is found in the central regions of some nearby galaxies (\citealt{Sandstrom2012}, in preparation). However, regardless of the environment, no significant radial trends in the virial mass-based \xco values are apparent in the central kiloparsecs of these galaxies. If anything, \xco may actually tend to decrease in NGC 6946 beyond a radius of 2~kpc. Note that CO (2-1) can be enhanced by slight increases in gas temperature and/or density \citep{Koda12} and is not as pure a tracer of gas surface density as CO (1-0). 

\section{Summary and Conclusions}

Using high resolution observations of the $^{12}$CO (J=1-0) transition, we map the bulk molecular gas in five substantial nearby spiral galaxies. Using the Clumpfind algorithm \citep{Williams94}, we resolve 182 individual GMCs within the central kiloparsecs of three galaxies; this is the largest such sample of clouds within galaxies analogous to the Milky Way. In addition, we detect GMCs within both the galactic centers and the inner disks of these galaxies, enabling comparisons of clouds in both environments with the corresponding cloud populations in the Milky Way. In general, we find good consistency between GMCs in the disk of the Milky Way as studied by S87 and GMCs in the central few kiloparsecs of the galaxies presented here even though we are only sensitive to analogs of the largest Galactic clouds due to our resolution limits. Specifically: \\
\indent $\bullet$ The sizes and velocity dispersions of the GMCs in nearby spirals are generally consistent with the trend exhibited by Milky Way disk GMCs. \\ 
\indent $\bullet$ Plotting GMC velocity dispersions against their CO luminosities also reveals behavior consistent with the Milky Way trend over the range in $\sigma_{v}$ in which Galactic GMCs are detected. The trend in CO luminosity and cloud size that we detect is indistinguishable from that seen in the Milky Way over more than four orders of magnitude in luminosity.\\ 
\indent $\bullet$ We do not detect a significant component of clouds with elevated velocity dispersions for a given size as observed by \citet{Oka01} at the Galactic Center. We find only a handful of such clouds, and these have luminosities consistent with the predictions from measurements of resolved GMCs in Local Group dwarfs and spirals (B08). \\
\indent $\bullet$ We find a linear relationship between GMC virial masses and CO luminosities, implying a constant value of \xco in each galaxy and supporting our assumption of virialization. The dominant source of the factor-of-two scatter between GMCs is intrinsic, leading to an average conversion factor in each galaxy within a factor of two of the value determined for the Milky Way at similar GMC masses. \\ 

No radial trends in \xco are apparent in the central kiloparsecs of each individual galaxy. If \xco is correlated with metallicity, even in the metal-rich centers of galaxies, then this finding is consistent with the shallow metallicity gradients in these regions and corresponding \xco predictions over the same areas \citep{Wilson95, Arimoto96}. In any event, our measurements help to anchor the metal-rich end of the purported relationship between metallicity and average conversion factor, which until now has relied largely on Galactic measurements. 

We estimate virial parameters for our clouds under the assumption that they are gravitationally bound (i.e., $\alpha_{vir}$ $\sim$ 1), as is supported by the linear relationship between cloud masses and luminosities. We calculate a very similar range of virial parameters for the Milky Way disk clouds in S87. GMCs in the galactic centers and inner disks, where physical conditions can vary, exhibit similar gravitational stability (as expressed by the virial parameter).

Comparing our population of clouds allows us to determine how appropriate the Galactic conversion factor typically assumed is for all such systems. The CO luminosities and virial masses measured for our resolved GMC sample indicate that, on average, \xco is within a factor of two of the typically assumed Galactic value of the conversion factor (2 \xcounits).  

\acknowledgments
JDM would like to thank the referee, Erik Rosolowsky, for his comments which have improved this manuscript, and Adam Leroy for helpful discussions about clumpfinding. This work is supported in part by the NSF under grant \# 1211680. JK also acknowledges support from NASA through grant NNX09AF40G, a Hubble Space Telescope grant, and a Herschel Space Observatory grant. This research has made use of the NASA/IPAC Extragalactic Database (NED) which is operated by the Jet Propulsion Laboratory, California Institute of Technology, under contract with the National Aeronautics and Space Administration. 

\bibliography{jen_refs}

\appendix
\begin{center}
COMPARISON OF CLUMPFIND AND {\it CPROPS} 
\end{center}

\renewcommand{\thefigure}{A-\arabic{figure}} 
\setcounter{figure}{0}  

The determination of GMC properties involves two major steps: (1) cloud identification in a 3-dimensional data cube, and (2) parameter measurements, including corrections for instrumental resolution. Cloud identification [step (1)] may be the single largest source of uncertainty in such analyses \citep{Rosolowsky06}. 

In this Appendix, we compare the cloud samples derived from the same data using two commonly used algorithms: Clumpfind \citep{Williams94} and \cprops \citep{Rosolowsky06}. We ultimately decide to adopt Clumpfind in this study for its treatment of step (1). Clumpfind is simpler than \cprops -- it has only two free parameters, while \cprops has in total 7 parameters (4 explicit plus 3 implicit, as described below) -- and, in any case, we find similar statistical results both qualitatively and quantitatively using both methods. As described in the text (\S 3 and \S 4.1), our treatment of step (2) is similar to that used in \cpropse.

In its treatment of step (1), the Clumpfind algorithm contours the data cube at multiples of its rms noise down to a user-specified minimum contour level, locates peaks within the contours, and follows them down to lower contours through connecting pixels. No clump profile is assumed. In our analysis, as described in \S 3, we examine the velocity profile of each clump and (conservatively) reject apparent blends, partial clouds, and clouds with significantly non-Gaussian velocity profiles. 

Alternately, the \cprops package first identifies potential clouds in three dimensions using boxes centered on local maxima. If more than one maximum exists within a box (the size of which is discussed below), the lesser maximum is rejected if they merge smoothly over their highest shared isocontour; this smooth merging is defined by comparing the unextrapolated moments associated with the two peaks. If none of the moments increase by more than 100\% for both maxima, no two of the moments increase by more than 50\% for both maxima, and the flux increases by less than 200\% for both maxima, then the clouds are considered to be smoothly merging, and the higher intensity maximum is kept while the lesser maximum is rejected. These three implicit parameters are used to pare maxima from the region \citep{Rosolowsky06}. Additionally, if the ratio between one peak and the highest shared contour between the two peaks is less than a value $\Delta T_{max}$, the lesser peak is rejected. If the maxima are found to be indistinct, both are rejected. In order to detect GMCs, the four (explicit) free parameters assumed are the size of the box (15~pc and 2~\kms~when running the algorithm with the defaults; otherwise these are scaled to the actual resolution), $\Delta T_{max}$ (2$\sigma_{rms}$, or 1~K in a 15~pc beam), and a parameter called T$_{clip}$ (at 2.5~K), which is used to smooth clumpy substructure within clouds and is therefore not particularly important for extragalactic GMCs. Running \cprops with an input data set where the channel size is greater than 2~\kms, the spatial resolution is greater than 15~pc, and the signal-to-noise is low (i.e., less than 10), as is the case with our observations (and with most other extragalactic GMC studies, including many which invoke \cpropse), is outside of the ``warrantied" parameter space where \cprops can be optimally run. In such cases, the authors caution that the decomposed results of the algorithm may not be directly comparable to other data sets.

To address step (2), \cprops deconvolves the beam size from the measurements in order to account for resolution biases and extrapolates the cloud sizes to $T$=0~K to account for signal-to-noise biases using a weighted, least squares fit to the higher cloud boundary temperature measurements of each cloud. In this study, we also deconvolve from our observations the Gaussian beam size of a point source with the same T$_{peak}$ as the clump, measured at the largest extent where the clump is measured (T=$\Delta$T) and unsmoothed velocity channel width, as recommended by both \citet{Williams94} and \citet{Rosolowsky06}. We calculate the three-dimensional sizes of our clouds using the measurement dispersions as described in \S 4.1 and in \citet{DonovanMeyer12}. Since our cloud parameters are derived from measurements out to a contour level twice the rms noise in each cube, not to an extrapolated $T$=0~K, the \citet{Williams94} equations [Equations 1 and 2] are appropriate for correcting our measurements for the resolution elements.

\begin{figure}
\plottwo{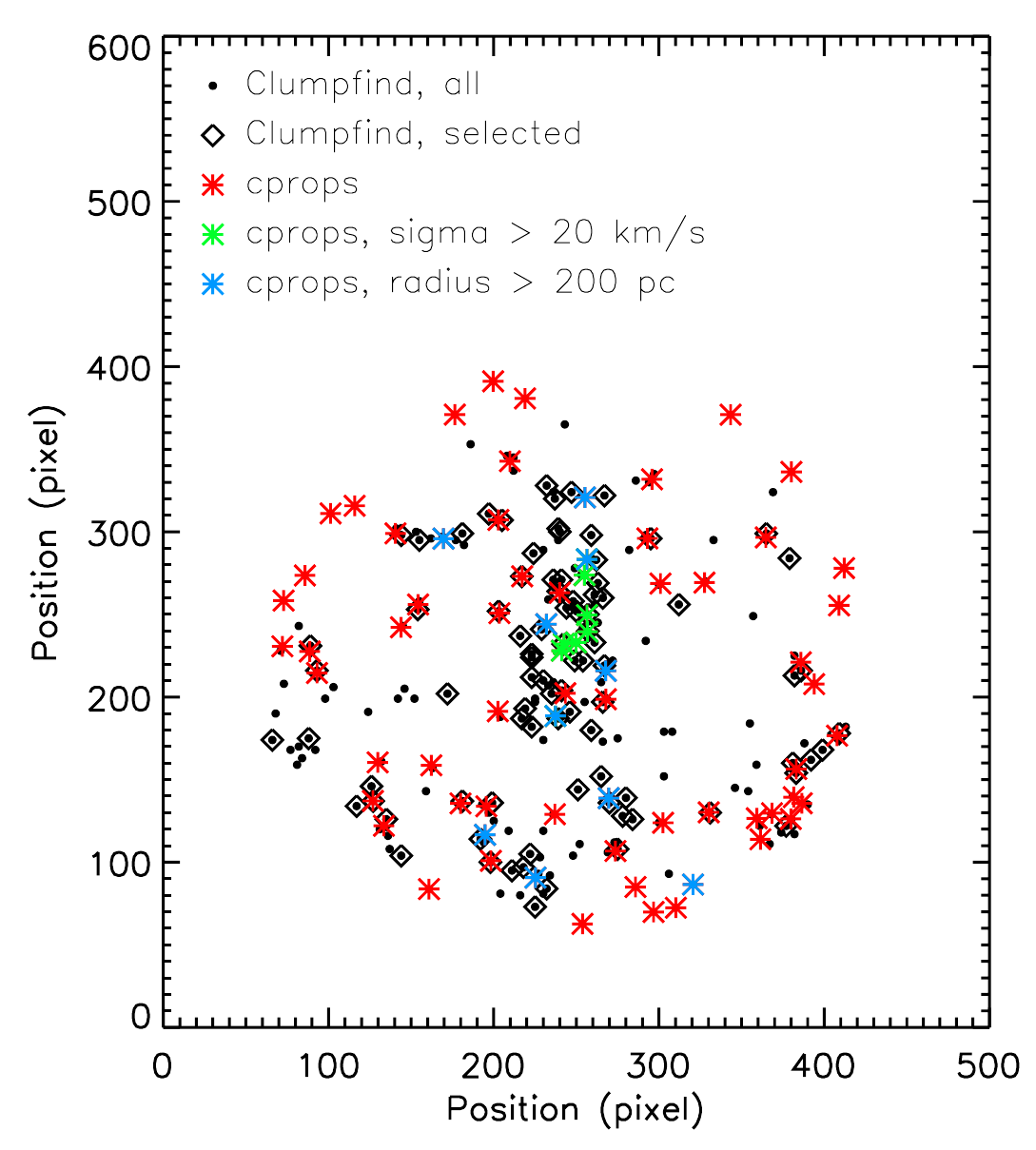}{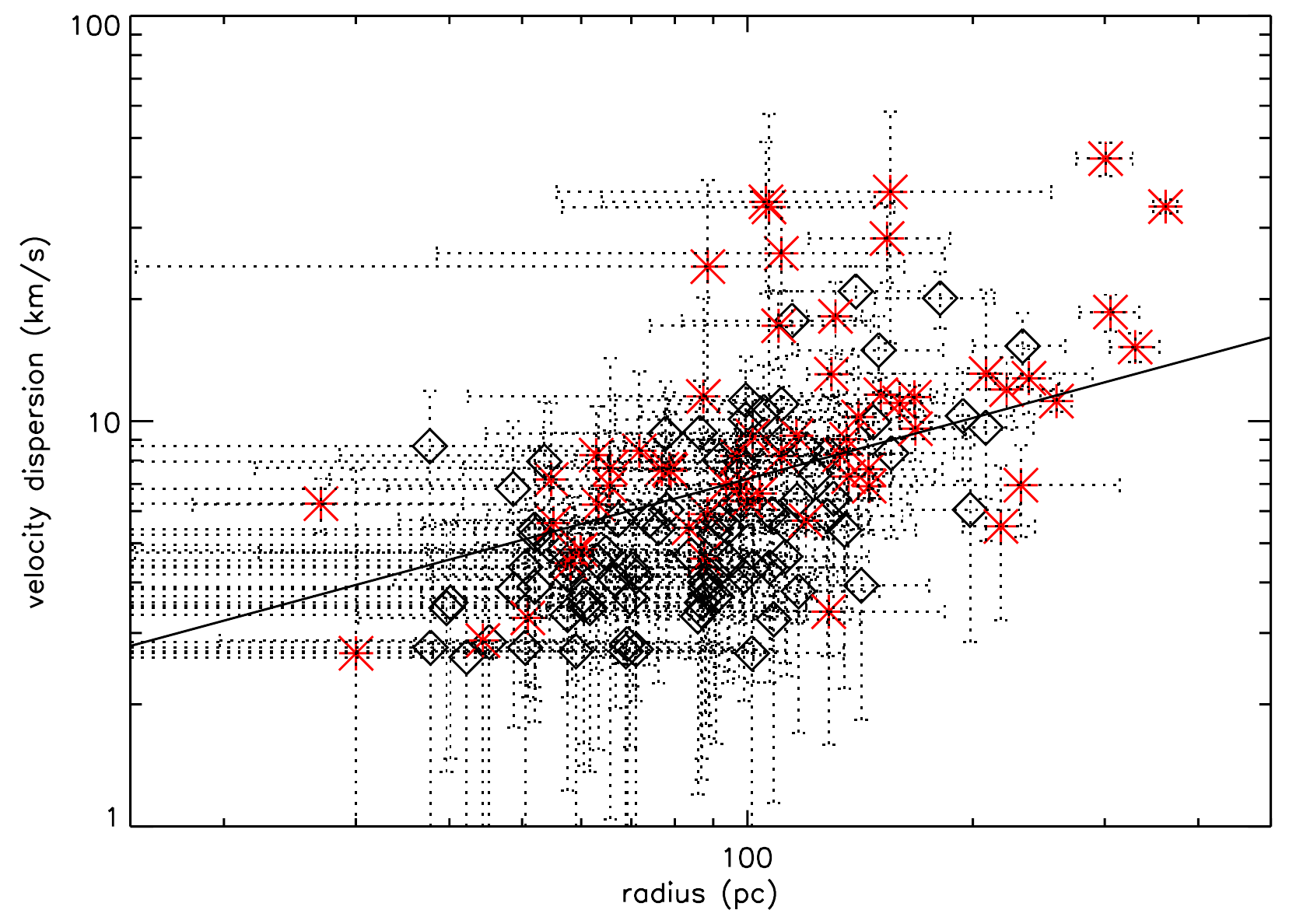}
\caption{$\it{Left:}$ Relative locations of the clouds found using Clumpfind and cprops. $\it{Right:}$ Radii, velocity dispersions, and associated errors of the clouds measured using both algorithms. The Galactic relationship found by S87 (0.72 $\sigma_{v}^{0.5}$) is also shown. \label{cprops1} }
\end{figure}

We compare the results for NGC 6946 using both algorithms, running Clumpfind with the parameters described in \S 3 and running \cprops with the default parameters. \cprops is designed to locate and measure cloud properties without regard to a specific cutoff in the signal-to-noise ratio, so the 2$\sigma$ rms noise limit that we designate in Clumpfind does not apply here. The resulting \cprops sample consists of 72 clouds, while the Clumpfind sample consists of 200 clouds (after our flux and resolution cuts) and 87 clouds after our velocity profile analysis. We show the relative locations of the clouds derived using both algorithms in Figure~\ref{cprops1}. Clumpfind deconvolves 70\% of the CANON cube emission into clouds; 65\% of the emission is located in clouds which survive the flux and resolution cuts, and 38\% of the emission is located in clouds which survive our velocity profile analysis, for a total mass of 5.8 $\times$ 10$^{8}$ \Msun~in GMCs. By comparison, \cprops deconvolves 22\% of the cube emission into clouds for a total mass of 2.5 $\times$ 10$^{9}$ \Msun~in GMCs (or 4.9 $\times$ 10$^{8}$ \Msun~if the deconvolved, non-extrapolated sizes and velocity dispersions are used to calculate cloud masses). Since Clumpfind contours all emission in the field and finds peaks within the contours, it allocates much more CO emission into clouds than \cprops does, and (mostly) through our velocity profile cuts, the percentage of emission detected in clouds is reduced. As should be expected, the amount of mass detected by \cprops using the unextrapolated measurements is more consistent with our measurement than the total mass in the extrapolated clouds; the extrapolation for cloud envelopes increases the total mass in \cprops clouds by a factor of 5.

Though fewer clouds are detected by \cprops overall compared to Clumpfind, for the most part, each \cprops cloud is near a corresponding Clumpfind cloud. In general, Clumpfind separates emission into smaller clumps that \cprops designates as a larger, single clump. Specifically in the central region, \cprops finds a few very large clouds where Clumpfind breaks the emission into a larger number of smaller clouds. We note that clumps with apparently blended velocity profiles are discarded by our visual inspection method, while \cprops may keep them as a single entity if they are found to be smoothly merging. 

Except for a handful of clouds with elevated radii and/or velocity dispersions (relative to the Clumpfind sample) detected by \cpropse, the trends between measured radii and velocity dispersions -- shown in the right panel of Figure~\ref{cprops1} -- of the two sets of cloud measurements are largely consistent with one another. The ranges of radii recovered for the clouds are similar even though \cprops performs an extrapolation to the 0~K contour to simulate perfect sensitivity. The S87 size-velocity dispersion relationship is overplotted on the measurements; most clouds in the two samples obtained using different algorithms, which define clouds in such different ways, are largely consistent with the Galactic relation. 

In the spiral arms and the galactic center, where more clumps are detected (and we are more likely to discard them for being blends), \cprops apportions emission into clouds with larger velocity dispersions and/or radii (Figure~\ref{cprops1}). Particularly in the center, \cprops tends to measure clouds with elevated velocity dispersions for their size (compared to the Clumpfind measurements and relative to the S87 relationship). If the corresponding emission in Clumpfind does not exhibit a regular velocity profile, such clumps would be rejected from our sample. Regions where \cprops finds clouds with larger sizes than the corresponding Clumpfind clouds tend to be in more crowded areas of spiral arms. In these cases, Clumpfind tends to break the emission comprising a single \cprops cloud with a large radius and/or velocity dispersion into multiple smaller clouds. 

We conclude that for our extragalactic sample, where our resolution is adequate to resolve the largest GMCs, the two algorithms produce results largely consistent with the Galactic GMC size-velocity dispersion relation. However, for our analysis, we prefer to present our results using the Clumpfind algorithm for its simpler treatment of clump deconvolution. We also employ a conservative definition of blends and are additionally more inclusive of clouds with near-Gaussian profiles and velocity dispersions near our instrumental resolution.

\end{document}